\documentclass[conference,compsoc]{IEEEtran}
\ifCLASSOPTIONcompsoc
  \usepackage[nocompress]{cite}
\else
  \usepackage{cite}
\fi
\ifCLASSINFOpdf
\else
\fi
\usepackage{sankey}
\usepackage{algorithm}
\usepackage{algpseudocode}
\usepackage{multirow}
\usepackage{booktabs}
\usepackage{colortbl}
\usepackage[margin=1in]{geometry}
\usepackage{xcolor}
\usepackage{amsmath}
\definecolor{lightgray}{gray}{0.9}
\definecolor{darkgray}{gray}{0.7}
\definecolor{lightgreen}{RGB}{200,255,200}
\definecolor{greenish}{RGB}{180,255,180}
\usepackage[table]{xcolor}
\usepackage{booktabs}
\usepackage{multirow}
\usepackage{graphicx}
\usepackage{rotating}
\usepackage{colortbl}
\usepackage{geometry}
\usepackage{array}
\usepackage{stfloats}
\usepackage{listings}
\usepackage{xcolor} 
\usepackage{enumitem}
\usetikzlibrary{arrows,automata,positioning}
\usepackage{pgfplots}
\pgfplotsset{compat=1.18}
\usepackage{tcolorbox}
\usepackage{amssymb}
\definecolor{darkgray}{gray}{0.75}
\definecolor{lightgray}{gray}{0.9}
\definecolor{highlight}{RGB}{180,255,180}
\renewcommand{\arraystretch}{1.25}
\definecolor{catrow}{RGB}{235,243,252}
\definecolor{darkgray}{gray}{0.75}
\definecolor{lightgray}{gray}{0.9}
\definecolor{highlight}{RGB}{180,255,180}
\definecolor{codegreen}{rgb}{0,0.6,0}
\definecolor{codegray}{rgb}{0.5,0.5,0.5}
\definecolor{codepurple}{rgb}{0.58,0,0.82}
\definecolor{backcolour}{rgb}{0.95,0.95,0.92}

\lstdefinestyle{mystyle}{
    backgroundcolor=\color{backcolour},   
    commentstyle=\color{codegreen},
    keywordstyle=\color{magenta},
    numberstyle=\tiny\color{codegray},
    stringstyle=\color{codepurple},
    basicstyle=\ttfamily\footnotesize,
    breakatwhitespace=false,         
    breaklines=true,                 
    captionpos=b,                    
    keepspaces=true,                 
    numbers=left,                    
    numbersep=5pt,                  
    showspaces=false,                
    showstringspaces=false,
    showtabs=false,                  
    tabsize=2
}
\usepackage{tabularx,booktabs,xcolor}
\usepackage{makecell}
\usepackage{multirow}
\usepackage{subcaption}
\usepackage{booktabs,xcolor}
\usepackage{xspace}
\usepackage{url} 
\newcommand{\commentout}[1]{}

\newcommand{\motif}{\textsc{MOTIF}\xspace}
\newcommand{\benchmark}{\textsc{MOTIF-bench}\xspace}

\newcommand{\agent}{\textsc{MOTIF-agent}\xspace}
\newcommand{\model}{\textsc{MOTIF-model}\xspace}

\newcommand{\eg}{e.g.\xspace}

\begin{document}
%
\title{Exposing Hidden Interfaces: LLM-Guided Type Inference for Reverse Engineering macOS Private Frameworks}

\author{\IEEEauthorblockN{Arina Kharlamova}
\IEEEauthorblockA{MBZUAI\\
Abu Dhabi, UAE\\
Arina.Kharlamova@mbzuai.ac.ae}
\and
\IEEEauthorblockN{Youcheng Sun}
\IEEEauthorblockA{MBZUAI\\
Abu Dhabi, UAE\\
Youcheng.Sun@mbzuai.ac.ae}
\and
\IEEEauthorblockN{Ting Yu}
\IEEEauthorblockA{MBZUAI\\
Abu Dhabi, UAE\\
Ting.Yu@mbzuai.ac.ae}
}


%


\maketitle

\begin{abstract}
Private macOS frameworks underpin critical services and daemons but remain undocumented and distributed only as stripped binaries, complicating security analysis. We present MOTIF, an agentic framework that integrates tool-augmented analysis with a finetuned large language model specialized for Objective-C type inference. The agent manages runtime metadata extraction, binary inspection, and constraint checking, while the model generates candidate method signatures that are validated and refined into compilable headers. On MOTIF-Bench, a benchmark built from public frameworks with ground-truth headers, MOTIF improves signature recovery from 15\% to 86\% compared to baseline static analysis tooling, with consistent gains in tool-use correctness and inference stability. Case studies on private frameworks show that reconstructed headers compile, link, and facilitate downstream security research and vulnerability studies. By transforming opaque binaries into analyzable interfaces, MOTIF establishes a scalable foundation for systematic auditing of macOS internals.
\end{abstract}

%
\IEEEpeerreviewmaketitle

\section{Introduction and Motivation}
\label{sec:introduction}

Modern software ecosystems increasingly depend on closed-source components and proprietary libraries embedded within operating systems. This is especially evident in Apple's macOS.  macOS ships with numerous private frameworks,  which are collections of code libraries that provide essential services but lack official documentation or support~\cite{Long_2024,private_2025}. Reverse engineering is often the only means for security analysts and developers to understand these Application Programming Interfaces (APIs), identify vulnerabilities, and enable third-party solutions. 
\commentout{
However, the task is notoriously difficult: compiled Mach-O binaries omit high-level type information, and manually reconstructing function signatures from assembly code is both error-prone and labor-intensive. The implications are significant, as undocumented APIs can conceal critical security flaws or impede interoperability with external software~\cite{Blazytko_2023,Caliskan_Yamaguchi_Dauber_Harang_Rieck_Greenstadt_Narayanan_2018,Romanov_Kurtukova_Fedotova_Shelupanov_2023}. Consequently, the field increasingly requires advanced, automated methodologies for reverse engineering~\cite{Wang_Wang_Wu_2015,Zhuo_Zhang_2025}.
}


\begin{figure}[h!] 
    \centering
    \includegraphics[width=\linewidth]{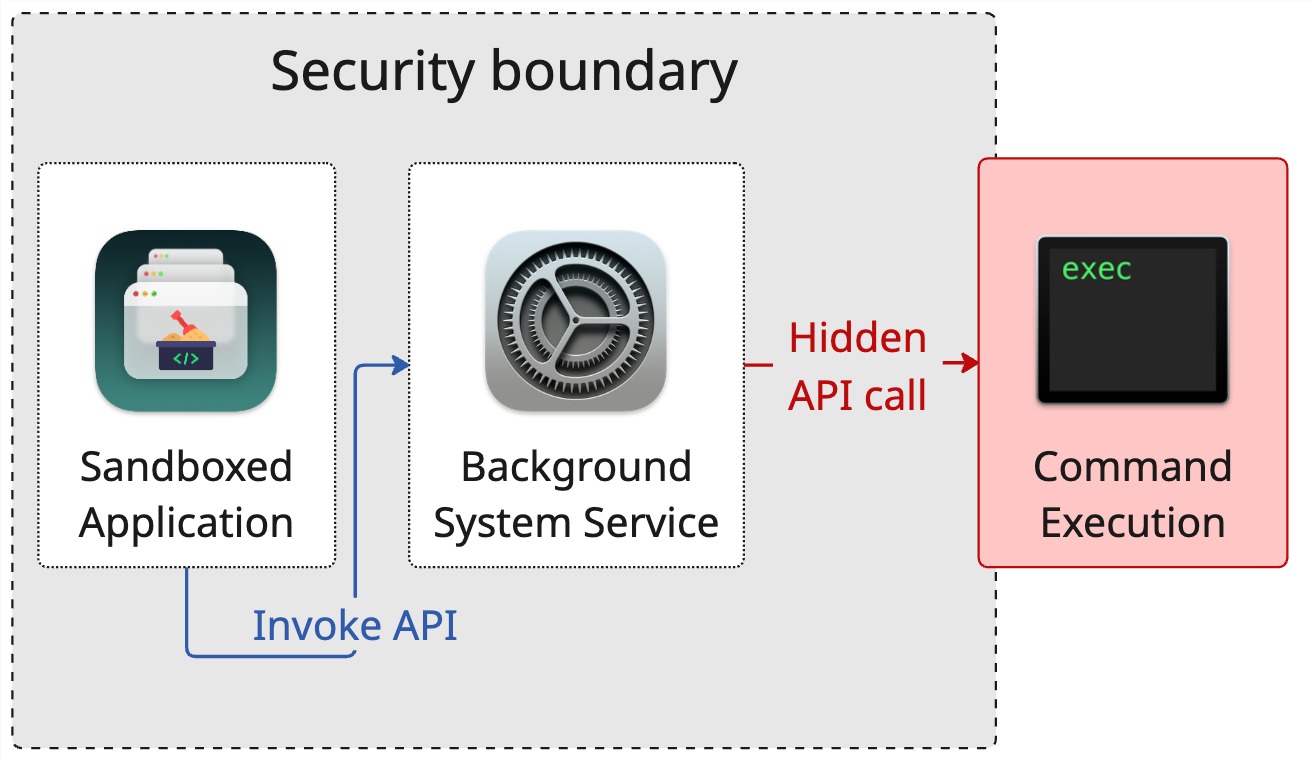}
    \caption{\emph{Command execution across privilege boundaries via hidden APIs.} 
    A sandboxed application invokes a private framework API, which registers an XPC messaging interface. Through this channel, attacker-controlled arguments can be used to execute code outside the sandbox.}
    \label{fig:introduction_example}
\end{figure}

Private macOS frameworks have repeatedly been at the center of severe security flaws~\cite{Blazytko_2023,Caliskan_Yamaguchi_Dauber_Harang_Rieck_Greenstadt_Narayanan_2018,Romanov_Kurtukova_Fedotova_Shelupanov_2023}. Consider a recent case involving the \textit{StorageKit} framework in macOS Sonoma beta. Simply loading this private framework caused the system to automatically register a background service for inter-process communication (an XPC service). Deeper inspection revealed an undocumented messaging interface that allowed sandboxed applications to execute arbitrary commands with attacker-controlled arguments. By inferring the structure of this hidden interface, researchers reconstructed a communication proxy and demonstrated remote command execution from a sandboxed context. Although removed in macOS 14.0, this case reflects a recurring pattern of vulnerabilities (illustrated in Figure~\ref{fig:introduction_example}): a hidden API exposes an unintended capability, which then becomes an entry point for privilege escalation or policy bypass. Earlier examples include CVE-2019-8561 \cite{nistCve20198561} in PackageKit, which enabled System Integrity Protection (SIP) bypass via a race condition, and CVE-2021-30873 \cite{nistCve202130873}, which permitted process injection through private interface misuse. Accurate type inference in such contexts helps surface overly broad interfaces and implicit trust boundaries, offering structural cues that can inform vulnerability analysis without requiring direct exploit development.

Automated extraction and completion of private framework APIs represent a key step in the broader reverse engineering process for macOS security analysis. Accurate inference of API argument and return types enables researchers to reconstruct interfaces and reason about their intended use, shifting the research focus from tedious manual recovery to higher-level vulnerability detection. This capability extends beyond macOS itself: many third-party applications (\eg, \cite{githubGitHubLwouisalttabmacos}, \cite{githubGitHubKoekeishiyayabai}), including widely distributed commercial software (\eg, \cite{folivoraFolivoraaiGreat}), rely on undocumented APIs for extended functionality. Vulnerabilities in these APIs can therefore propagate into higher-level software that inherits or misuses them. Improved access to reconstructed API definitions (Type inference) produced through type inference thus benefits both OS-level security research and the auditing of dependent applications.

Recovering type information from binary executables has long been a core objective in reverse engineering. Classical approaches rely on static analysis with handcrafted rules or data-flow reasoning to infer variable and function types from usage patterns~\cite{Katz,pei}. For example, TIE~\cite{Tie} introduced a systematic reconstruction framework that applies constraint propagation across binary code to improve both precision and conservativeness compared to traditional decompiler heuristics. Subsequent work explored machine learning for identifying type signatures in disassembled code~\cite{smith}. Notably, TypeMiner~\cite{maier2019typeminer} employed classification techniques to recover data types, achieving 76–93\% accuracy on C binaries. 

Despite these advances, significant challenges remain: existing methods often struggle with complex object-oriented binaries, depend on large training corpora, or fall short in reconstructing complete method signatures and interface definitions. As a consequence, in practice, reverse engineers still expend significant manual effort to determine how undocumented macOS frameworks should be invoked~\cite{exploring-macos-private-frameworks.html_2025,macOSPrivateFrameworks}. Persistent limitations hinder prior work: rule-based systems are brittle against evolving compiler idioms, while purely static analyses remain incomplete without broader semantic understanding. Existing learning-based efforts focus on general binary analysis rather than macOS framework reconstruction, leaving this domain largely unexplored.

Meanwhile, although LLMs have shown strong performance in source-code understanding and reasoning~\cite{alebachew2025aiguidedexplorationlargescalecodebases, 10795058, zhuo2025ccprobabilisticllmmethods}, their application to binary analysis remains largely unexplored. This motivates our central question: \emph{Can LLMs, when combined with program-analysis tools, autonomously recover accurate method signatures and interface definitions in undocumented macOS frameworks?} To address this, we introduce \textsc{MOTIF (Mach-O Type Inference Framework)}, a system that treats type inference and interface reconstruction as an iterative, LLM-guided analytical process. \motif tackles key challenges in binary analysis: \textsc{(I)} sparse type encodings, \textsc{(II)} incomplete disassembly metadata, and \textsc{(III)} the absence of ground-truth evaluation through three components:
\begin{enumerate}
    \item \agent: An LLM-guided reverse engineering pipeline that orchestrates external tools such as disassemblers, decompilers, and static linters to progressively reconstruct missing type information from incomplete headers (see \S\ref{sec:motif}). 
    \item \benchmark: A benchmark dataset constructed from public macOS frameworks with ground-truth headers, used for quantitative evaluation of type inference accuracy (see \S\ref{sec:motifbench}). 
    \item \model: A small LLM distilled from large-model interaction traces, specialized for local deployment and efficient type inference (see \S\ref{sec:motifmodel}).
\end{enumerate}

\section{Related Work and Comparative Context}
\label{sec:related-work}



\textbf{Community Reconstructions of Private Frameworks.}
Technical blogs and curated indexes~\cite{jviotti2023,bunn2019} (\eg, \texttt{PrivateFrameworks} \cite{githubPrivateFrameworks} repositories and The Apple Wiki \cite{theapplewikiAppleWiki}) document private APIs by extracting Objective-C metadata with tools such as \texttt{class-dump} and \texttt{RuntimeBrowser}. These sources are valuable as living catalogs and as starting points for auditing undocumented interfaces. However, the reconstructed headers are typically incomplete: many parameter and return types are recovered only as \texttt{id} or \texttt{void*}, which encode minimal semantic information and hinder static analysis or safe API usage. Additionally, signatures involving blocks (i.e., function-type closures) and protocol-qualified types (interfaces that constrain object behavior) are often missing or incorrectly inferred.

\textbf{Type Reconstruction in Compiled Binaries.}
Constraint-based and dataflow-driven systems for stripped binaries (\eg, ~\cite{bosamiya3trex, noonan2016polymorphictypeinferencemachine, Tie, 299717, maier2019typeminer, 2664, 9741274, 10.1145/3519939.3523449} ) infer types from calling conventions, register usage, and memory layouts. This line of work is strong in statically typed C/C++ settings with regular Application Binary Interfaces (ABIs) and compiler artifacts. Objective-C breaks these assumptions in several fundamental ways. 

First, message sends store only compact type encodings for selector names (\eg, \texttt{[obj doSomething:]}) but omit type annotations, making it difficult to determine parameter and return types from the call site alone. 

Second, dynamic dispatch defers method resolution until runtime, shifting essential interface information into Objective-C–specific binary sections (\eg, \texttt{\_\_objc\_classlist}, 
\texttt{\_\_objc\_methname}), which fall outside the scope of conventional static analysis pipelines. 

Third, the widespread use of \texttt{id} (a dynamically typed object placeholder) further erases static constraints: any object can be passed or returned under the \texttt{id} type, which carries no information about supported methods or internal structure. 

Finally, blocks (closures capturing local context) introduce additional complexity through nested function signatures and implicit state capture. Classical propagation techniques, which rely on explicit type flow and fixed call graphs, cope poorly with such semantics. As a result, they often fail to recover protocol-qualified types or infer generic collection types accurately.

\textbf{Large Language Models in Binary Analysis.}
Recent studies apply LLMs to reverse engineering tasks, with a growing number of works exploring static and dynamic contexts~\cite{Liu_2025, Tan_2024, shang2025binmetriccomprehensivebinaryanalysis, pordanesh2024exploringefficacylargelanguage, chen2025recopilotreverseengineeringcopilot, 10.1145/3728958}. ReSym~\cite{10.1145/3658644.3670340} recovers variable and data-structure symbols from stripped binaries, while PentestGPT~\cite{299699} explores autonomous offensive workflows. These approaches primarily target C or C++ binaries and are not designed for the dynamic, metadata-rich environment of macOS private frameworks. They generally operate independently of external tools such as disassemblers, metadata extractors, or static/dynamic linters, producing results that may appear plausible but cannot be systematically verified against the binary.

\section{Background: macOS Frameworks and Objective-C Typing}
\textbf{Mach-O Binaries and Framework Architecture.} Apple’s macOS and iOS use the Mach-O (Mach Object) file format as the native binary format for executables, libraries, and dynamically-loaded components. 

In macOS, shared libraries are often packaged as frameworks, which are directory bundles (with a \texttt{.framework} extension) containing a dynamic shared library along with resources like headers, assets, and metadata. A framework bundle typically includes a versioned directory structure (\eg, \texttt{Versions/A}) that houses the actual Mach-O library and symlinks at the top level pointing to the current version. This design allows multiple versions of a library to coexist for binary compatibility, with the loader automatically linking against the current version of the framework’s \texttt{dylib}. 

Public macOS frameworks (located in \texttt{/System/Library/Frameworks/}) export stable APIs with accompanying header files and documentation. In contrast, private frameworks (in \texttt{/System/Library/PrivateFrameworks/}) are intended for internal OS use and are not documented or exposed in the official SDK. These private frameworks are structurally similar to public ones (they are Mach-O dynamic libraries with Objective-C or C/C++ code inside), but because they lack published headers or documentation, their APIs are opaque to third-party developers. System applications and daemons frequently rely on private frameworks for functionality; for example, Disk Utility links against numerous private frameworks like \texttt{DiskManagement.framework }and \texttt{StorageKit.framework}. 

Although private frameworks are not part of the official SDK, they still expose exported classes and selectors that serve as their callable interface. These symbols must remain accessible for any system or third-party binary that links against the framework, even though no 
documentation or header files are provided. 

Finally, because macOS frameworks (including private ones) often make heavy use of Objective-C, their binaries embed Objective-C metadata (class names, method selectors, etc.) in Mach-O sections. This runtime metadata, along with the Mach-O structure, is key for disassemblers and other tools to interpret the contents of frameworks. In summary, macOS private frameworks are Mach-O dynamic libraries packaged in bundle form, loaded via the dynamic linker from a shared cache in modern systems. They are functionally analogous to public frameworks but without the benefit of documentation or readily available symbol information, which poses a significant challenge for analysis and type recovery.

\textbf{Headers and Their Role in Reverse Engineering.} Header files (C/C++ header files or Objective-C interface files) describe the public interface of libraries by declaring functions, methods, classes, constants, and data types. In the context of macOS frameworks, headers define the API contract (\eg for example, for a public framework like \texttt{Foundation} or \texttt{AppKit}), Apple provides \texttt{.h} files in the Xcode SDK that list all class definitions, method signatures, and data structures developers can use. These headers are invaluable for both compilation and for understanding what functions exist and how to call them. 

Naturally, no such headers are provided for private frameworks, which means reverse engineers must reconstruct the APIs themselves. To mitigate this, researchers use header reconstruction tools. One well-known utility is \texttt{class-dump}, which analyzes a Mach-O binary and generates Objective-C interface declarations (pseudo-headers) by reading the Objective-C runtime metadata embedded in the file. For example, running \texttt{class-dump} on a private framework binary will produce a header file with the class declarations and method signatures that the binary contains. This is how early iOS/macOS enthusiasts uncovered hidden APIs: by dumping private frameworks to see what classes and methods Apple has implemented. In a similar vein, the open-source \texttt{RuntimeBrowser} tool on macOS can load all private frameworks and use Objective-C runtime APIs to enumerate classes and selectors, presenting a list of methods and allowing export of header files. This approach bypasses direct Mach-O parsing but still extracts only the encoded selectors embedded in the binaries. Disassemblers like \texttt{Hopper} and \texttt{IDA Pro} also have features or plugins to export headers (\eg they leverage the same Objective-C metadata to reconstruct class interfaces). 

However, these reconstructed headers have limitations. \textsc{(I)} \emph{Missing symbols.} If a binary has been stripped of names or heavily optimized, the extracted output will be incomplete. \textsc{(II)} \emph{Lack of type context.} The generated headers often show generic placeholders (\texttt{id}, \texttt{void*}) because the extracted metadata contains only selector names and encoded type stubs, without information about actual usage. Without analyzing how methods are called (\eg, whether an \texttt{id} is consistently a \texttt{NSString*}), these tools cannot reconstruct precise types. \textsc{(III)} \emph{Ambiguous signatures.} As a result, many method signatures look cryptic or truncated, with little information about the real classes or data structures involved. \textsc{(IV)} \emph{Non-ObjC code is invisible.} Pure C functions and C++ methods (which may exist in the same framework) are not exposed in Objective-C metadata, so they must be identified through disassembly or other forms of analysis.

\textbf{Objective-C Selectors and Message Passing.} A large portion of macOS private frameworks are implemented in Objective-C, making an understanding of its dispatch model essential. Objective-C is a dynamically dispatched, message-passing language: rather than invoking methods through fixed function pointers or vtable offsets, it sends messages to objects. 

At compile time, a call such as \texttt{[object doSomething:arg]} is lowered into a call to the C function \texttt{objc\_msgSend}, which takes the receiver, a selector (\texttt{SEL}) identifying the method name, and the call arguments. 

At runtime, the Objective-C system resolves the selector by inspecting the receiver’s class and consulting its method dispatch table, climbing the inheritance chain if necessary. This process realizes late binding, with the actual target implementation chosen dynamically based on the object’s class. 

At the binary level, this indirection leaves only partial traces. The compiled code issues calls to \texttt{objc\_msgSend}, while selector strings and their method encodings are stored in dedicated sections such as \texttt{\_\_objc\_methname} and \texttt{\_\_objc\_methtype}. A method encoding compactly describes the return type and argument layout (\eg, \texttt{c44@0:8@16c24Q28\^\{\}@36}), but the actual method body is not directly referenced. For reverse engineering, this means one can readily observe which selector is invoked, yet resolving the implementation requires inferring the receiver’s class. 

In practice, without static type information, most Objective-C call sites in disassembly collapse into the same generic \texttt{objc\_msgSend}, which makes it difficult to reconstruct call relationships and reason about object behavior. This indirection highlights why accurate type inference in Objective-C binaries is inherently challenging.


\section{Scope, Assumptions, and Threat Model}

We consider two related scenarios: direct inspection of a private framework binary, and analysis of a client application that depends on a private framework and thus provides an entry point or usage context. Here, framework denotes the particular private framework under investigation. Dependencies from a client binary can be discovered by \textsc{(I)} inspecting the application's \texttt{Info.plist} and entitlement entries (for example, Mach service names, entitlement flags, or declared access to hardware/services), which often indicate which system services or private frameworks the binary interacts with, or \textsc{(II)} using dynamic-analysis commands (\eg, \texttt{lm} in \texttt{lldb}) to list linked libraries at runtime. 

The targeted private framework satisfies several common properties:  \textsc{(I)} \emph{Objective-C implementation.} The framework is written in Objective-C and therefore exposes selector names as metadata. \textsc{(II)} \emph{Partial API surface.} It exposes callable interfaces to dependent system components or applications, but without distributed headers describing exact argument and return types. This absence of type definitions prevents analysts from recovering full method signatures directly from the binaries. These conditions reflect the broader challenge: private frameworks are discoverable and usable, but lack the type signatures necessary for rigorous analysis. 


In this scenario, an attacker with standard reverse-engineering capabilities (\eg, able to statically inspect binaries and dynamically load them into controlled environments, but without access to source code) has two primary strategies for dealing with the framework: \textsc{(I)} Decompile a client binary that depends on the framework (\eg, a system application such as Mail, Calendar, or Disk Utility, or a third-party app that links the private framework). \textsc{(II)} Directly reverse engineer the framework itself, reconstructing pseudo-headers with class and method declarations. While this provides structural visibility, the exact method types for arguments and return values remain unresolved. It is precisely this gap that our type-inference framework aims to address.

\section{\agent: LLM-guided Reverse Engineering framework}
\label{sec:motif}

\begin{figure}[t!]
    \centering
    \includegraphics[width=0.5\textwidth]{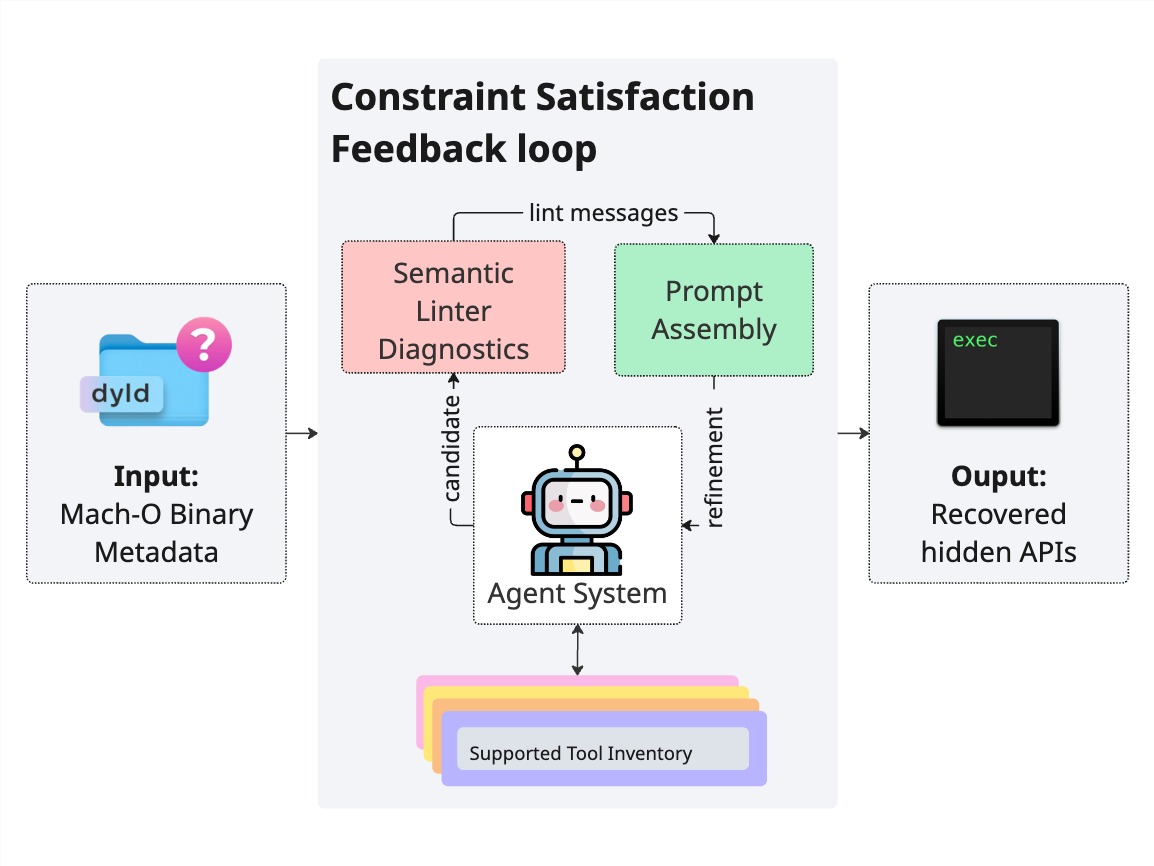}
     \caption{\emph{Example of \motif in operation:} the agent interacts with analysis tools and a semantic linter in a closed-loop refinement process to recover hidden APIs.}
    \label{fig:framework-example}
\end{figure}

\begin{figure*}[t!]
    \centering
    \includegraphics[width=\textwidth]{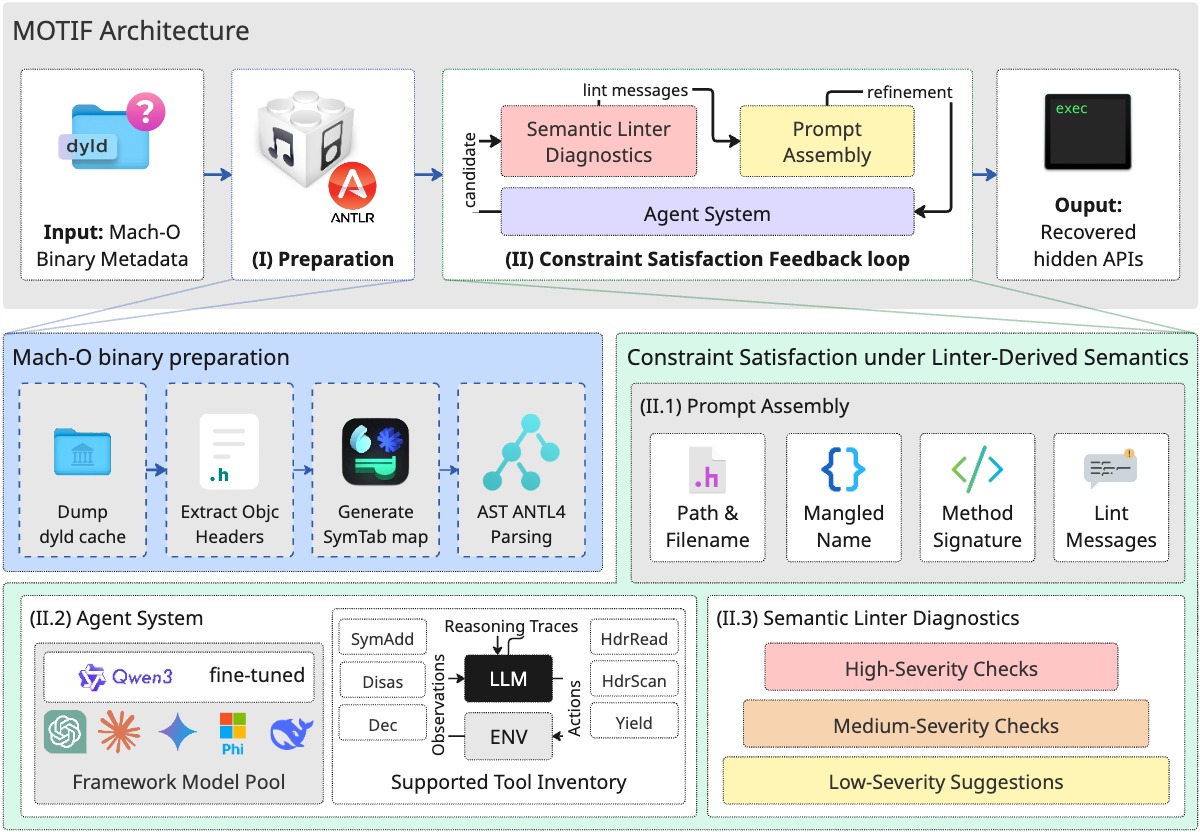}
     \caption{\emph{Overview of the \motif architecture.} The pipeline consists of: (I) binary preparation; (II) a constraint satisfaction feedback loop under linter-derived semantics, composed of (II.1) prompt assembly; (II.2) an agent system; and (II.3) semantic linter diagnostics. The loop iteratively refines candidate signatures until constraint satisfaction is achieved.}
    \label{fig:framework-scheme}
\end{figure*}


\agent combines LLM reasoning with static constraints to recover Objective-C method signatures from stripped macOS binaries (Figure~\ref{fig:framework-example}). An end-to-end inference trace appears in Appendix~\ref{sec:appendix_example}.

\textbf{From Embedding Priors to Constraint-Guided Inference.} Modern LLMs exhibit emergent competence in specialized domains even without task-specific supervision \cite{berti2025emergentabilitieslargelanguage}, \cite{wei2022emergentabilitieslargelanguage}. \textit{Our premise is that Objective-C idioms, naming conventions, and method patterns, in particular  from Apple SDKs, are sufficiently represented in pretraining corpora} (\eg, developer blogs, \texttt{StackOverflow} answers, public headers) to form strong embedding priors: latent associations encoded in its vector representations that link semantically or syntatically related concepts (\eg methods named \texttt{initWithCoder:} usually take an argument of type \texttt{NSCoder *}). However, embeddings alone are insufficient for recovering nontrivial type signatures in private frameworks. The absence of header files, presence of anonymous structs, and stripped symbols create gaps that require grounded reasoning beyond what the model can recall. These limitations motivate a hybrid architecture that combines LLM-driven inference with constraint-guided feedback. The next section outlines the high-level design of this system, detailing how preparation, prompting, tool usage, and refinement interact to recover hidden types from stripped binaries.

\textbf{High-Level Architecture.} Figure~\ref{fig:framework-scheme} illustrates the full architecture of \motif, which operates as a hybrid inference loop interleaving pretrained LLM priors with symbolic and structural constraints derived from static analysis. The architecture decomposes into two main stages: (\textsc{I}) binary preparation and (\textsc{II}) constraint-guided inference.

\textbf{Stage I: Static Binary Preparation (Inputs and Parsing).} The system accepts as input a stripped Mach-O binary and extracts partial metadata, specifically Objective-C headers and symbol maps, using \texttt{ipsw}. Extracted headers are then parsed with a customized ANTLR4-based parser to identify underspecified declarations and candidate methods requiring type recovery.

\textbf{Stage II: Constraint-Guided Inference (Context, Tools, Linter, Loop).} This context is packaged into structured prompts that combine:
\begin{itemize}
    \item method string with underspecified types,
    \item mangled symbol name,
    \item tool definitions (available static and dynamic analysis primitives),
    \item metadata (\eg, file path, framework name).
\end{itemize}

During inference, the model agent can query these tools within a ReAct loop, retrieving disassembly, typedef lookups, or symbol addresses as needed. Candidate Objective-C method signatures are synthesized and then validated by a semantic source-level linter enforcing structural, idiomatic, and compilation-level constraints. Diagnostic messages from the linter are fed back into the prompt construction step, creating a closed-loop refinement system. The loop terminates when all hard constraints specified in the system configuration are satisfied or a fixed number of refinement iterations is reached.

\textbf{Iterative Constrained Refinement.} Unlike static prompting, our system performs \textit{iterative constrained refinement} by integrating the LLM within a tightly controlled feedback loop. Each model-generated signature is statically verified through a domain-specific linter developed in this work for macOS and Objective-C binaries, which emits structured warnings and hard constraints (\eg, unresolved anonymous structs, syntactically invalid pointer types, unsafe generics such as \texttt{NSDictionary<NSString, id>}).

\textbf{Interactive Diagnostic Cycle and Convergence.} The model is then re-prompted with this feedback, forming an \emph{interactive diagnostic cycle}. This constrained loop acts as a \textit{syntactic and semantic convergence mechanism}, filtering out implausible completions and driving the model toward signatures that satisfy Objective-C compiler requirements, idiomatic conventions, and disassembler-aligned structures.

\textbf{Constraint Taxonomy and Linter Semantics.} Our semantic linter emits structured diagnostics over candidate signatures, capturing violations of Objective-C typing and private/public framework conventions. Table~\ref{tab:linter-taxonomy} summarizes the taxonomy.

\begin{table*}[t!]
\caption{Constraint taxonomy enforced by the linter during type refinement. Each constraint defines a pattern to match over partial or complete method signatures and emits structured messages with severity levels. High-severity constraints block generation; medium-severity ones act as soft guidance.}
\scriptsize
\setlength{\tabcolsep}{6pt}
\renewcommand{\arraystretch}{1.3}
\begin{tabular}{l l l p{4.5cm} p{4.5cm}}
\toprule
\textbf{Name} & \textbf{Severity} & \textbf{Message Type} & \cellcolor{orange!10}\textbf{Example (Violation)} & \cellcolor{green!10}\textbf{Suggested Correction} \\
\midrule

\rowcolor{catrow}
\multicolumn{5}{l}{\textit{Syntactic and Structural Constraints (High Severity)}} \\
SyntaxErrors & High & Syntax Violation & \texttt{void) doSomething;} & \texttt{(void) doSomething;} \\
NoStructs & High & Inline Struct Detected & \texttt{struct \{ double x0; double x1; \} center} & \texttt{CGPoint center} \\
SelectorMismatch & High & Selector Divergence & \texttt{doSomething:argument2:} & \texttt{doSomething:withArg2:} \\
\midrule
\rowcolor{catrow}
\multicolumn{5}{l}{\textit{Semantic Typing Constraints (Medium Severity)}} \\
StructRefs & Medium & Raw Struct Pointer Used & \texttt{struct \_NSZone *zone} & \texttt{NSZoneRef zone} \\
GenericCollections & Medium & Missing Generic Parameter & \texttt{NSArray args} & \texttt{NSArray<NSString *> args} \\
NoIdGenerics & Medium & Generic Type is \texttt{id} & \texttt{NSArray<id> values} & \texttt{NSArray values} \\
\midrule
\rowcolor{catrow}
\multicolumn{5}{l}{\textit{Semantic Typing Constraints (Low Severity)}} \\
ConventionalTypes & Low & Non-conventional Scalar Type & \texttt{\_Bool isEnabled;} & \texttt{BOOL isEnabled;} \\

\bottomrule
\end{tabular}
\label{tab:linter-taxonomy}
\end{table*}

\begin{figure}[b]
    \centering
    \includegraphics[width=\linewidth]{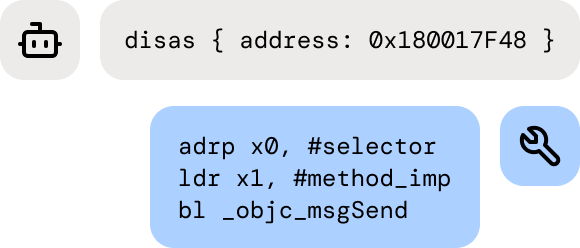}
    \caption{Example tool call: LLM invokes disassembler to inspect method body.}
    \label{fig:tool-example}
\end{figure}

\textbf{Tool Interface and Execution Layer.} As formalized in Appendix~\ref{sec:formalization}, inference targets are derived from partial headers extracted from Mach-O binaries, parsed into abstract syntax trees, and filtered according to underspecification criteria. In Figure~\ref{fig:tool-example}, we illustrate a real interaction where the model queries the disassembler for address \texttt{0x180017F48}, observes the use of \texttt{objc\_msgSend}, and revises its candidate method type accordingly.

\textbf{Tool-Augmented Inference.} Once target is constructed, the pipeline enters the inference phase. Here, the LLM operates as a tool-augmented agent, iteratively invokes tools, integrates returned signals, and converges toward a type-complete candidate signature satisfying both syntactic and semantic constraints. This loop forms the core of our type recovery mechanism.

\textbf{Tool Inventory.} The following tools are currently exposed to the model:

\begin{itemize}[leftmargin=4em]
    \item[$\texttt{SymAddr}$] \emph{Symbol Address Resolution.}
    Maps a selector to its corresponding memory address via a static lookup over symbol tables extracted from the firmware image using \texttt{ipsw}.
    
    \item[$\texttt{Disas}$] \emph{Disassembly View.}  
    Given an address, returns a disassembled instruction trace (ARM64) from a fixed-length window. Type inferences are drawn from operand behavior (\eg, pointer dereferences, integer arithmetic, or usage of \texttt{objc\_msgSend} targets).
    
    \item[$\texttt{Dec}$] \emph{Decompiler View.}  
    Invokes a decompiler backend \texttt{IDA Pro CLI} to emit higher-level pseudocode. While incomplete for Objective-C, this often reveals control flow, return-type hints (\eg, scalar vs. reference), and selector dispatches on argument slots.
    
    \item[$\texttt{HdrRead}$] \emph{Header Inspection.}  
    Retrieves the full header in which the current selector is declared, enabling access to nearby type definitions, property declarations, and superclasses.
    
    \item[$\texttt{HdrScan}$] \emph{Header Index Scan.}  
    Enumerates the available Objective-C headers for the current framework bundle. This enables resolution of unknown types (\eg, \texttt{CustomViewModel}) to their defining interface.
    
    \item[$\texttt{Yield}$] \emph{Terminalization Operator.}  
    Produces the final inferred method signature and exits the ReAct loop. This tool is required to safely conclude inference in instruction-tuned models trained for aggressive tool-calling.
    
\end{itemize}

Each tool is embedded into the agent’s action space and selected via token-level planning during decoding. Tool outputs may be validated downstream by the semantic linter or recycled via constrained retries in the Iterative Constrained Refinement.
\section{\benchmark: A Reproducible Benchmark for Mach-O Type Inference}
\label{sec:motifbench}

\begin{figure*}[b!] 
    \centering
    \includegraphics[width=\textwidth]{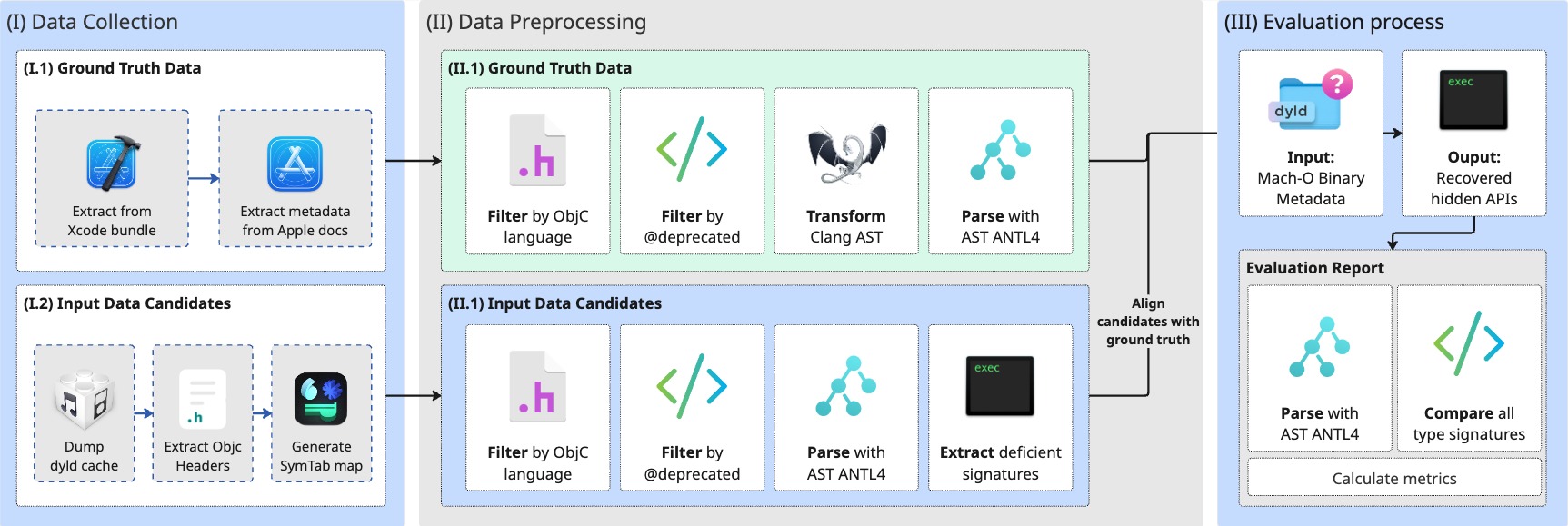}
    \caption{\textbf{Overview of Mach-O Type-Inference Framework Benchmark construction process.} The pipeline consists of five main stages. \textbf{Data collection:} Ground truth data are extracted from Xcode resources, while input data candidates are dumped from the dyld shared cache and processed. \textbf{Filtering:} Frameworks are binned by method count and equally sampled to ensure balanced representation across varying sizes and prevent sampling bias. \textbf{Data preprocessing:} Extracted headers are transformed and parsed using ANTLR4-based tools to reduce size and extract incomplete method signatures. \textbf{Matching:} Candidate methods are matched against the ground truth and unmatched methods are discarded. \textbf{Evaluation process:} Solutions generated by LLMs are parsed and compared against the ground truth on a per-signature basis, evaluating argument types and return types to generate benchmark metrics and reports.}
    \label{fig:wide}
\end{figure*}

We introduce the first benchmark specifically designed to evaluate the ability of analysis tools and language models to recover type signatures from compiled macOS frameworks. To our knowledge, no prior benchmark targets this problem space: existing datasets for code completion or decompilation do not capture the challenges unique to Objective-C binaries, where dynamic message passing and method encodings complicate static recovery. 

Constructing such a benchmark is non-trivial, as it requires aligning authoritative ground-truth headers with incomplete binary metadata while ensuring balanced coverage across framework categories and scales. Motivated by the need for a principled evaluation environment to measure framework-level type inference performance, we developed \benchmark. The benchmark is constructed from publicly documented Apple frameworks and incorporates real binary metadata, yielding a version-specific, reproducible, and objective foundation for type-inference evaluation. 

Beyond serving as an evaluation suite, the benchmark and its datasets can also be used to fine-tune inference models (see Section~\ref{sec:motifmodel}), enabling a consistent pipeline from supervised training to empirical assessment.

\begin{figure}[ht!] 
    \centering
    \includegraphics[width=\linewidth]{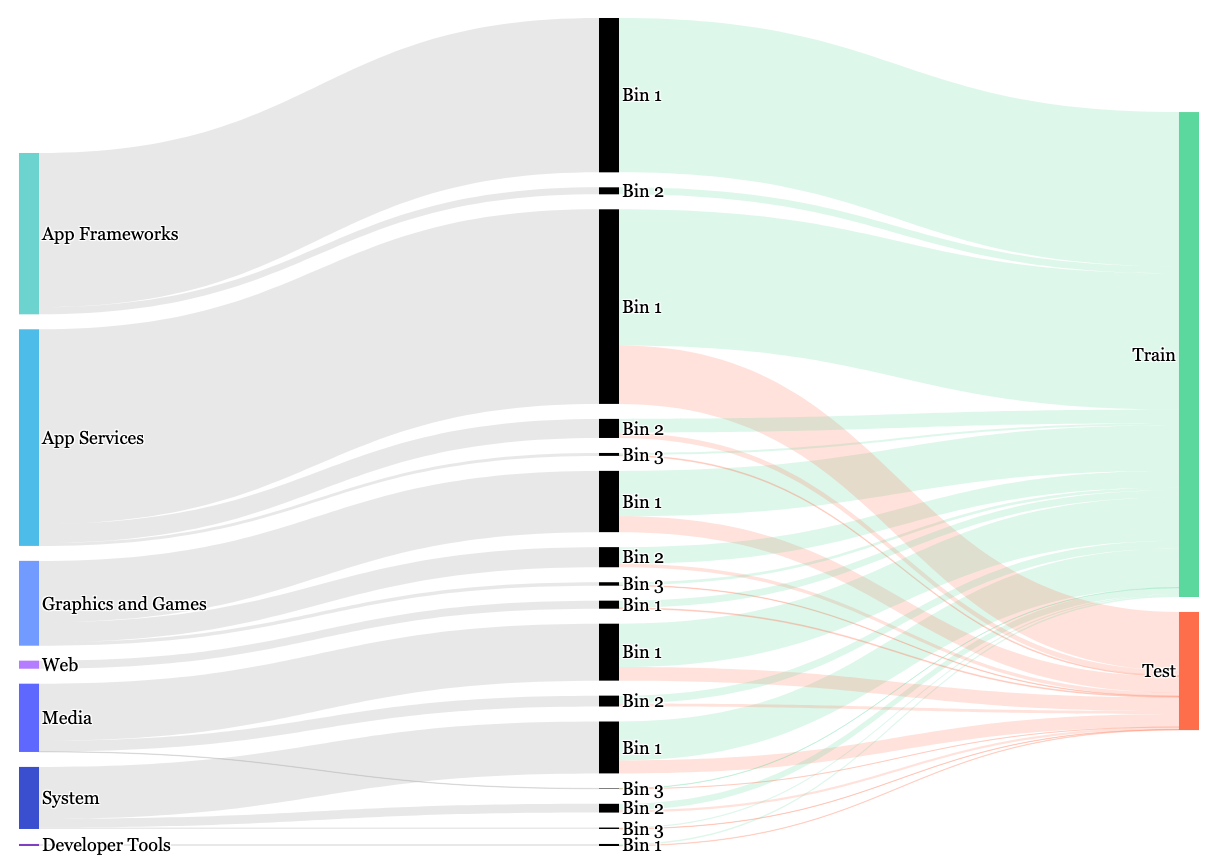}
    \caption{Example of dataset generated from MacOS Tahoe 26.0 Beta (25A5295e)}
    \label{fig:dataset_example}
\end{figure}

\textbf{Benchmark Construction.} To provide a fair and reproducible evaluation environment, we constructed a dynamic dataset pipeline that extracts macOS frameworks directly from the target system version. The pipeline collects two complementary sources: \textsc{(I)} ground-truth headers from the corresponding version of Xcode, and \textsc{(II)} binary metadata from the system's \texttt{dyld} shared cache. This dual-source design ensures that benchmark instances reflect both authoritative type specifications and the incomplete artifacts available in practice. The specific set of frameworks included varies with the macOS version under analysis; for example, the distribution shown in Figure~\ref{fig:dataset_example} corresponds to macOS 26.0 Beta (build 25A5295e). Frameworks are stratified by Apple’s official category labels (\eg, System, Graphics \& Games, App Services). Within each category, frameworks are further partitioned into bins by method count: small ($\leq 10$), medium (10–100), and large (100–1000). From each bin, frameworks are sampled in equal proportion, mitigating sampling bias and preventing large frameworks from dominating the dataset. Approximately 70\% of sampled frameworks are reserved for evaluation, with the remainder allocated to model training dialogues. The detailed allocation is summarized in Table~\ref{tab:train_test_ratios}.

\begin{table}[h]
\caption{Train/Test Ratios Across Categories}
\centering
\begin{tabular}{lcc}
\hline
\textbf{Category} & \textbf{Train Ratio} & \textbf{Test Ratio} \\
\hline
System & 0.74 & 0.26 \\
Graphics and Games & 0.76 & 0.24 \\
App Services & 0.73 & 0.27 \\
Developer Tools & 0.77 & 0.23 \\
Media & 0.74 & 0.26 \\
App Frameworks & 1.00 & 0.00 \\
Web & 0.92 & 0.08 \\
\hline
\textbf{Total} & \textbf{0.81} & \textbf{0.19} \\
\hline
\end{tabular}
\label{tab:train_test_ratios}
\end{table}

This construction procedure yields a benchmark that is both balanced and extensible. By grounding in official Apple headers and dyld caches, \benchmark captures the practical difficulty of type recovery while maintaining a clear ground truth. Its dynamic, version-specific design ensures that the dataset can be regenerated for future macOS releases, supporting long-term reproducibility of evaluations.

\section{\model: Lightweight Tool-Aware Language Model for Type Inference}
\label{sec:motifmodel}
Reverse engineering private or undocumented macOS frameworks often involves legally sensitive binaries, making reliance on proprietary API-based systems undesirable due to confidentiality, compliance, or data protection concerns. Furthermore, cloud APIs introduce high token costs and unpredictable latency, complicating reproducibility in controlled research settings. A lightweight, locally deployable model provides a privacy-preserving alternative. From a technical perspective, general-purpose instruction-tuned LLMs are not designed to synthesize type signatures under partial information, nor are they optimized to integrate tool-derived constraints. \model addresses this gap. Its core design goal is to recover Objective-C method signatures when headers are absent, while conforming to structural and symbolic constraints available from static analysis. To support this objective, the model is explicitly aligned with tool-aware inference: it must not hallucinate signatures, but rather integrate tool outputs when available and abstain or defer when constraints cannot be satisfied.  

\textbf{Training Dataset and Preprocessing.} The training corpus was distilled from \benchmark, ensuring alignment between evaluation and fine-tuning. Approximately 3,000 dialogues were collected by running multiple candidate LLMs on benchmark tasks. To ensure quality, we applied a distillation filter with the following criteria: \textsc{(I)} \emph{Malformed tool invocations} were excluded to avoid training on invalid call sequences. \textsc{(II)} \emph{Infinite recursion in the ReAct loop} was discarded, preventing degenerate reasoning traces. \textsc{(III)} \emph{Low-quality generations} with benchmark scores below 0.8 were removed.  

The remaining dialogues captured complementary strengths across models, for example, some systems were more effective at generic collection typing, while others were better at resolving struct references. By merging them, we obtained a single coherent dataset of high-quality dialogues. We applied the Qwen-3 chat template to the dataset, converting all dialogues into a consistent multi-turn format with explicit tool calls and definitions preserved. Preprocessing enforced a structured prompt schema, where tool-related context was marked using control tokens such as \texttt{<tools>} for tool definitions and \texttt{<tool\_call>} for tool invocations. This ensured that the model could reliably distinguish natural-language dialogue turns from tool-driven actions.   


\textbf{Fine-Tuning and Tool-Constraint Alignment.} Fine-tuning was performed using the \textsc{QLoRA} approach~\cite{qlora} via the Axolotl framework~\cite{axolotl}. \textsc{QLoRA} (Quantized Low-Rank Adaptation) operates on quantized weight representations while inserting low-rank adapters into frozen transformer layers, enabling efficient adaptation with a reduced memory footprint. This setup allowed training to complete on a single NVIDIA H200 GPU. Constraint alignment was realized by embedding tool definitions into the prompt context and penalizing generations that violated linter-enforced rules. In effect, the model was trained to synthesize type signatures consistent with tool-derivable evidence, rather than relying solely on prior knowledge.  

\textbf{Inference-Time Tool Integration.} At runtime, \model operates within a ReAct loop, where it must conform to available tool traces and abstain or flag inconsistencies if conflicts arise. Early experiments revealed a tendency toward aggressive tool invocation, sometimes exhausting the loop without converging on a stable signature. To mitigate this, we introduced a control operator that allows the model to explicitly exit the loop and return to the feedback cycle, which stabilized long-horizon reasoning and prevented unproductive tool churn. The final model was quantized and converted into Apple’s MLX format, enabling efficient execution on a single M2 Ultra (32 GB VRAM) or comparable consumer-grade accelerators. This deployment footprint allows researchers to conduct local, privacy-preserving type inference experiments at scale, without exposing binaries or research traces to cloud providers.

\section{Evaluation Methodology and Results}
\label{sec:experiment-setup}
\begin{table*}[t!]
\caption{ Evaluation on MOTIF-Bench. \textbf{Dark gray} indicates best among all models, \textbf{light gray} best among open-source. Green cells mark top-3 ranks within each group. \textbf{Avg.} denotes the average over all Signature Inference Subtasks (\textit{BT Completion}, \textit{Collection Inference}, \textit{DP Inference}, \textit{BT Inference}). \textbf{Rank} is computed based PM acc.
}
\scriptsize
\setlength{\tabcolsep}{2.5pt}
\begin{tabular}{rr|c|c|cccccc|cccc}
 & & & &
\rotatebox{60}{PM Accuracy} & \rotatebox{60}{EM Accuracy} & \rotatebox{60}{Tool Usage Rate} & \rotatebox{60}{Inference Stability} & \rotatebox{60}{Tool-Call Correctness} &
\rotatebox{60}{Tool-Call HR} & \rotatebox{60}{BT Completion} & \rotatebox{60}{Collection Inference} & \rotatebox{60}{DP Inference} & \rotatebox{60}{BT Inference} \\
\multicolumn{2}{r|}{\textbf{Methods}} & \textbf{Rank} & \textbf{Avg.} &
\multicolumn{6}{r|}{\cellcolor{orange!10}\textbf{Numerical / Binary Answer Metrics}} &
\multicolumn{4}{r}{\cellcolor{yellow!10}\textbf{Signature Inference Subtasks}} \\
\midrule

\rowcolor{catrow}
\multicolumn{14}{l}{\textit{Baseline}} \\
Static Analysis & & 17 & 24.5 & 14.9 & 0.0 & -- & -- & -- & -- & 98.0 & 0.0 & 0.0 & 0.0 \\
\midrule

\rowcolor{catrow}
\multicolumn{14}{l}{\textit{One-Shot No-Tooling LLMs}} \\
Claude Sonnet 4 & & 6 & 78.7 & 80.1 & 62.8 & -- & -- & -- & -- & 99.0 & \cellcolor{lightgray} 65.9 & \cellcolor{lightgray} 73.8 & \cellcolor{lightgray} 76.0 \\
Deepseek R1  & & 13 & 48.1 & 58.9 & 43.6 & -- & -- & -- & -- & 71.7 & 47.9 & 36.9 & 39.4 \\
Gemini 2.0 flash  & & 10 & 57.3 & 71.5 & 48.6 & -- & -- & -- & -- & 100.0 & 39.6 & 43.1 & 46.6 \\
ChatGPT 4o  & & 8 & 63.0 & 74.5 & 51.8 & -- & -- & -- & -- & 100.0 & 50.7 & 52.3 & 49.0 \\
\midrule

\rowcolor{catrow}
\multicolumn{14}{l}{\textit{Ours: Task-Aligned Tool-Calling LLMs (finetuned)}} \\
Qwen3 (8B) finetuned & & \cellcolor{highlight!20} 3 & 42.4 & \cellcolor{lightgray} 83.5 & \cellcolor{lightgray} 64.7 & \cellcolor{lightgray} 99.5 & 16.3 & 100.0 & 0.0 & \cellcolor{lightgray} 99.4 & 32.2 & 9.6 & 28.6 \\
Llama 3.1 (8B-Instruct) finetuned  & & 6 & 37.6 & 80.3 & 58.1 & 97.9 & 9.1 & 100.0 & 0.0 & 93.6 & 33.0 & 3.8 & 20.1 \\
\midrule

\rowcolor{catrow}
\multicolumn{14}{l}{\textit{Proprietary Models (API)}} \\
GPT-4o & & \cellcolor{highlight!50} 2 & 50.0 & 86.3 & 68.9 & \cellcolor{darkgray} 99.9 & 14.4 & 100.0 & 0.0 & 99.4 & 44.3 & 17.0 & 39.2 \\
Deepseek R1  & & 7 & 54.8 & 79.5 & 61.1 & \cellcolor{darkgray} 99.9 & \cellcolor{darkgray} 53.7 & 97.7 & 2.3 & 96.0 & 42.0 & 35.8 & 45.5 \\
Claude Sonnet 4 & & \cellcolor{highlight} 1 & 94.7 & \cellcolor{darkgray} 86.7 & 72.3 & 77.4 & 0.1 & \cellcolor{darkgray} 100.0 & 0.0 & 100.0 & \cellcolor{darkgray} 83.2 & \cellcolor{darkgray} 100.0 & \cellcolor{darkgray} 95.5 \\
Gemini-2.0 Flash & & 4 & 88.7 & 82.9 & 60.8 & 66.8 & 6.5 & 100.0 & 0.0 & \cellcolor{darkgray} 100.0 & 79.2 & \cellcolor{darkgray} 100.0 & 75.5 \\
XAI Grok 4 & & 5 & 56.4 & 82.1 & \cellcolor{darkgray} 73.0 & 99.7 & 5.6 & 100.0 & 0.0 & 82.1 & 50.7 & 48.1 & 44.6 \\
\midrule

\rowcolor{catrow}
\multicolumn{14}{l}{\textit{Open-source Models}} \\
Qwen3 (14B) & & 9 & 39.1 & 73.8 & 49.5 & 97.7 & \cellcolor{lightgray} 46.6 & 100.0 & 0.0 & \cellcolor{lightgray} 99.4 & 31.4 & 1.9 & 23.8 \\
gpt-oss-120b & & 11 & 41.5 & 68.3 & 61.6 & 85.2 & 40.9 & 98.3 & 1.7 & 81.1 & 31.1 & 23.8 & 29.9 \\
gpt-oss-20b & & 12 & 28.5 & 57.1 & 40.2 & 82.4 & 30.4 & 82.6 & 17.4 & 68.8 & 16.7 & 0.0 & 28.6 \\
Llama 3.1 8B Instruct & & 14 & 19.4 & 35.9 & 19.0 & 91.5 & 10.3 & 100.0 & 0.0 & 58.5 & 7.6 & 2.0 & 9.4 \\
Phi-3-medium & & 13 & 26.8 & 46.1 & 22.7 & 0.0 & 22.6 & -- & -- & 82.2 & 9.1 & 3.8 & 11.9 \\
\bottomrule
\end{tabular}
\label{tab:motifbenchresults}
\end{table*}

Our evaluation assesses both the empirical accuracy and practical utility of \textsc{MOTIF}. We \textsc{(I)} \emph{quantitative benchmark} on public macOS frameworks using \textsc{MOTIF-Bench} to measure type-inference accuracy against ground-truth headers; and \textsc{(II)} \emph{qualitative validate} on \textsc{MOTIF-Private}, a manually-curated dataset of private frameworks designed to assess reconstruction fidelity and practical utility in real-world reverse-engineering scenarios.

\textbf{Experimental Environment.} All experiments were conducted on a dedicated workstation running macOS \texttt{Sonoma 14.7.6 (23H626)} with an Apple M2 Ultra processor and 192 GB of unified memory.

\textbf{Toolchain.} The binary extraction pipeline relied on \texttt{ipsw} for dyld shared cache parsing and framework extraction, integrated features of \texttt{ipsw} for disassembly, and \texttt{IDA Pro 9.1} for decompilation. 
Our custom \texttt{ANTLR4}-based grammar was used to parse incomplete Objective-C headers into abstract syntax trees (ASTs) and identify inference targets. 
Static validation was performed with a bespoke type-consistency linter. 

\textbf{Models.} The primary system under evaluation is \model, our fine-tuned 8B-parameter LLM based on Qwen3-8B and trained for tool-guided type inference. We compare this model against both proprietary API-based systems and open-source baselines. For comparison studies, we additionally evaluate one-shot prompting of these models without tool access, as well as a traditional static approach such as \texttt{ipsw}.

\textbf{Public Framework Benchmarking.} We begin by evaluating \motif on public macOS frameworks, where official headers provide reliable ground truth. All evaluation data are drawn from \benchmark, which contains stratified samples of public frameworks extracted from the dyld shared cache and paired with header files from the corresponding Xcode SDK release. The benchmark ensures balanced coverage across framework categories and method scales, reducing sampling bias that could otherwise distort accuracy measurements.

\textbf{Evaluation Metrics.} We evaluate type inference quality using three primary correctness metrics:  
\begin{enumerate}
    \item \emph{Partial Match (PM) Accuracy:} the average fraction of correctly inferred types across all argument and return positions. 
    Formally,
    \[
    \text{PM} = \frac{1}{|\mathcal{M}|} \sum_{m \in \mathcal{M}} \frac{|\{\, i \mid \hat{t}_{m,i} = t_{m,i} \,\}|}{|\{\, i \mid t_{m,i} \neq \bot \,\}|},
    \]
    where $\mathcal{M}$ is the set of evaluated methods, $t_{m,i}$ is the ground-truth type, and $\hat{t}_{m,i}$ is the inferred type. 
    This metric accounts for partial correctness, \eg, when some but not all types in a signature are recovered.  

    \item \emph{Exact Match (EM) Accuracy:} the fraction of methods for which the entire inferred signature (all argument and return types) exactly matches the ground truth. 
    This is a strict all-or-nothing measure: 
    \[
    \text{EM} = \frac{|\{\, m \in \mathcal{M} \mid \forall i: \hat{t}_{m,i} = t_{m,i} \,\}|}{|\mathcal{M}|}.
    \]
\end{enumerate}

Beyond correctness, we track behavioral metrics:  

\begin{enumerate}
    \setcounter{enumi}{3}
    \item \emph{Tool Usage Rate:} the proportion of inference tasks where the agent invokes at least one external analysis tool (disassembler, decompiler, or static linter). 
    This quantifies reliance on external program-analysis evidence rather than purely language-model predictions.  

    \item \emph{Inference Stability:} the proportion of methods for which the iterative inference process converges to a fixed signature within $K=10$ steps. 
    A method $m$ is considered stable if $\hat{t}_{m,i}^{(k)} = \hat{t}_{m,i}^{(k-1)}$ for all $i$ once some $k^{*} \leq K$ is reached.

    \item \emph{Tool-Call Correctness (TCC):} the fraction of tool invocations issued by MOTIF that exactly match the expected API usage (correct tool name, arguments, and file paths). 
    Each tool call is parsed and validated against a formal schema of permissible arguments. 
    TCC measures whether the agent is not only deciding to invoke a tool, but doing so in a syntactically and semantically valid manner.  

    \item \emph{Tool-Call Hallucination Rate (HR):} the proportion of tool invocations that are invalid, redundant, or unsupported (\eg, calls to nonexistent binaries, malformed flags, or references to artifacts not produced by prior steps). 
    This metric captures the extent to which the LLM attempts to fabricate functionality outside the actual tool API. 
\end{enumerate}

\begin{table*}[b!]
\scriptsize
\caption{\emph{Overview of Selected Private macOS Frameworks.} Each entry summarizes the framework’s prior security exposure (if any), brief functional role, and the specific usage contexts within the macOS ecosystem. Frameworks are grouped into three categories reflecting their source of partial ground truth, external tool dependencies, or breadth of application coverage.}
\setlength{\tabcolsep}{6pt}
\renewcommand{\arraystretch}{1.3}
\begin{tabularx}{\textwidth}{l l l X}
\toprule
\textbf{Private Framework} & \textbf{VulnHistory} & \textbf{\#} & \cellcolor{orange!10}\textbf{Short description} \\
\midrule
\rowcolor{catrow}
\multicolumn{4}{l}{\textit{Partial Annotations from Apple Open-Source}} \\
SafariFoundation & -- & 1 & Provides foundational services for Safari, including autocomplete, credential, and account handling across macOS. \\
DiskManagement & ~\cite{nistCVE20070117, nistCVE202440828} & 5 & Implements low-level disk volume management routines exposed to utilities and system daemons. \\
\midrule
\rowcolor{catrow}
\multicolumn{4}{l}{\textit{Community Reliance for Tooling}} \\
StoreFoundation & -- & 5 & Core framework supporting the Mac App Store’s app listing, download, update, and purchase operations. \\
CommerceKit & -- & 3 & Provides transactional and purchase APIs tied to Apple’s commerce and payment infrastructure. \\
\midrule
\rowcolor{catrow}
\multicolumn{4}{l}{\textit{Multi-Domain Coverage}} \\
DFRFoundation & -- & 1 & Framework for the Touch Bar subsystem, offering APIs to simulate and communicate with the Touch Bar.  \\
PIP & -- & 1 & Provides Picture-in-Picture support for video playback across macOS applications. \\
SystemUIPlugin & -- & 1 & Hosts extensions and menu bar plug-ins for system UI elements. \\
AOSKit & -- & 3 & Implements Apple Online Services authentication, including legacy SSO mechanisms. \\
SidecarCore & -- & 3 & Provides the core logic for Sidecar, enabling an iPad to function as a secondary display. \\
CalendarFoundation & ~\cite{theevilbitTALKExploiting} & 2 & Implements calendar event storage and scheduling logic underpinning Calendar.app. \\
DoNotDisturb & -- & 3 & Exposes APIs to control the Do Not Disturb (DND) system setting. \\
IMCore & -- & 5 & Provides messaging core for iMessage and SMS relay integration on macOS. \\
\bottomrule
\end{tabularx}
\label{tab:private-frameworks}
\end{table*}

\textbf{Signature Inference Subtasks-Level Metrics.} To capture fine-grained aspects of Objective-C type recovery, we evaluate MOTIF on four subtask-specific metrics. 
Each metric is defined over a subset of argument and return positions in the benchmark corpus $\mathcal{M}$, and accuracy is measured as the fraction of correctly recovered types within that subset. 

\begin{enumerate}
    \item \emph{Basic Type Completion (BTC):}  
    Let $\mathcal{S}$ be the set of positions whose ground-truth type is a primitive scalar (\eg, \texttt{int}, \texttt{BOOL}, \texttt{long}).  
    We compute
    \[
    \text{BTC} = \frac{|\{(m,i) \in \mathcal{S} \mid \hat{t}_{m,i} = t_{m,i}\}|}{|\mathcal{S}|}.
    \]

    \item \emph{Collection Inference (CI):}  
    Let $\mathcal{C}$ be the set of positions where the ground-truth type is a concrete Objective-C collection (\eg, \texttt{NSArray*}, \texttt{NSDictionary*}, \texttt{NSSet*}).  
    Accuracy is defined as
    \[
    \text{CI} = \frac{|\{(m,i) \in \mathcal{C} \mid \hat{t}_{m,i} = t_{m,i}\}|}{|\mathcal{C}|}.
    \]

    \item \emph{Delegate Protocol Inference (DPI):}  
    Let $\mathcal{P}$ be the set of positions where the ground-truth type includes a protocol-qualified annotation (\eg, \texttt{id<NSCopying>}).  
    We measure
    \[
    \text{DPI} = \frac{|\{(m,i) \in \mathcal{P} \mid \hat{t}_{m,i} = t_{m,i}\}|}{|\mathcal{P}|}.
    \]

    \item \emph{Block Type Inference (BTI):}  
    Let $\mathcal{B}$ be the set of positions whose ground-truth type is an Objective-C block, represented as a function pointer with explicit return and parameter types.  
    Equality requires the predicted block signature to match the ground truth in both return type and ordered parameter list.  
    Formally,
    \[
    \text{BTI} = \frac{|\{(m,i) \in \mathcal{B} \mid \hat{t}_{m,i} = t_{m,i}\}|}{|\mathcal{B}|}.
    \]
\end{enumerate}

\textbf{Baselines.} We evaluate MOTIF against two classes of baselines that capture the current state of practice in reverse engineering and automated type inference:  

\begin{enumerate}
    \item \emph{Metadata-based utilities.}  
    Tools such as \texttt{ipsw} reconstruct Objective-C headers directly from runtime metadata embedded in Mach-O binaries. This utility is widely used in both academic and practitioner settings but is limited to exposing selector names and generic placeholder types (\eg, \texttt{id}, \texttt{void*}), without recovering precise argument or return types.

    \item \emph{One-shot No-Tooling LLMs.}  
    We query general-purpose LLMs on the same incomplete method signatures but without tool access or iterative refinement. This baseline measures the extent to which improvements in MOTIF stem from tool integration and fine-tuning rather than raw model capability.  
\end{enumerate}

\subsection{MOTIF-Private: A Manually-Curated Private Framework Benchmark}
\label{sec:privcase}
\begin{table*}[t!]
\caption{\emph{Evaluation of Type Inference Accuracy on Private macOS Frameworks.}
We compare static analysis, our finetuned task-aligned tool-calling LLM (Qwen3 8B), and proprietary APIs across a diverse suite of numerical and structural inference metrics.}
\scriptsize
\setlength{\tabcolsep}{2.5pt}
\begin{tabularx}{\textwidth}{X|c|c|cccccc|cccc}
 & & &
\rotatebox{60}{PM Accuracy} & \rotatebox{60}{EM Accuracy} & \rotatebox{60}{Tool Usage Rate} & \rotatebox{60}{Inference Stability} & \rotatebox{60}{Tool-Call Correctness} &
\rotatebox{60}{Tool-Call HR} & \rotatebox{60}{BT Completion} & \rotatebox{60}{Collection Inference} & \rotatebox{60}{DP Inference} & \rotatebox{60}{BT Inference} \\
\multicolumn{1}{r|}{\textbf{Methods}} & \textbf{Rank} & \textbf{Avg.} &
\multicolumn{6}{r|}{\cellcolor{orange!10}\textbf{Numerical / Binary Answer Metrics}} &
\multicolumn{4}{r}{\cellcolor{yellow!10}\textbf{Signature Inference Subtasks}} \\
\midrule

\rowcolor{catrow}
\multicolumn{13}{l}{\textit{Baseline}} \\
Static Analysis & 9 & -- & 22.3 & 0.0 & -- & -- & -- & -- & \cellcolor{darkgray} 100.0 & 0.0 & 0.0 & 0.0 \\
\midrule

\rowcolor{catrow}
\multicolumn{13}{l}{\textit{Ours: Task-Aligned Tool-Calling LLMs (finetuned)}} \\
Qwen3 (8B) & \cellcolor{highlight} 1 & -- & \cellcolor{darkgray} 75.2 & \cellcolor{lightgray} 48.4 & 100.0 & 0.0 & 100.0 & 0.0 & \cellcolor{lightgray} 60.0 & \cellcolor{darkgray} 40.0 & 0.0 & \cellcolor{lightgray} 33.3 \\
\midrule

\rowcolor{catrow}
\multicolumn{13}{l}{\textit{Proprietary Models within MOTIF Framework (API Access)}} \\
GPT-4o & \cellcolor{highlight!20} 3 & -- & 71.9 & \cellcolor{lightgray} 51.6 & 100.0 & 0.0 & 100.0 & 0.0 & \cellcolor{lightgray} 60.0 & 0.0 & 0.0 & 33.3 \\
Deepseek R1 & \cellcolor{highlight!50} 2 & -- & 72.5 & 38.7 & 100.0 & \cellcolor{darkgray} 38.7 & 97.3 & 2.7 & \cellcolor{lightgray} 60.0 & 20.0 & 0.0 & 16.7 \\
Claude Sonnet 4 & 5 & -- & 66.7 &  \cellcolor{darkgray} 56.2 & 100.0 & 0.0 & 100.0 & 0.0 & 33.3 & 0.0 & 0.0 & \cellcolor{darkgray} 50.0 \\
Gemini-2.0 Flash & 4 & -- & 67.7 & 32.3 & 100.0 & 9.7 & 100.0 & 0.0 & \cellcolor{lightgray} 60.0 & 20.0 & 0.0 & \cellcolor{darkgray} 50.0 \\
XAI Grok 4 & 6 & -- & 58.3 & 48.4 & 100.0 & 6.5 & 100.0 & 0.0 & 40.0 & 0.0 & 0.0 & 0.0 \\
\midrule
\rowcolor{catrow}
\multicolumn{13}{l}{\textit{Open Source Models within MOTIF Framework (API Access)}} \\
Llama 3.1 8B Instruct  & 8 & -- & 24.6 & 6.5 & 83.9 & 0.0 & 100.0 & 0.0 & 40.0 & 0.0 & 0.0 & 20.0 \\
Phi-3-medium & 7 & -- & \cellcolor{lightgray} 37.3 & \cellcolor{lightgray} 6.5 & 0.0 & 6.5 & -- & -- & \cellcolor{lightgray} 60.0 & 0.0 & 0.0 & 0.0 \\
\bottomrule
\end{tabularx}
\label{tab:benchresults}
\end{table*}
While quantitative benchmarking provides measurable accuracy against public frameworks with known headers, the true motivation of our method lies in reverse engineering private macOS frameworks. To assess practical utility in this setting, we also manually constructed a small dataset of case studies across a curated set of private frameworks. Framework selection followed targeted criteria rather than random sampling, focusing on methods and frameworks that met at least one of the following conditions: \textsc{(I)} availability of partial ground truth through Apple’s open-source repositories; \textsc{(II)} active reliance by the community for building tools that are otherwise impossible without private frameworks; \textsc{(III)} historical association with vulnerabilities and security advisories; and \textsc{(IV)} coverage of diverse domains. An overview of these selected frameworks, including their security exposure (if any), functional role, and usage contexts within the macOS ecosystem, is provided in Table~\ref{tab:private-frameworks}. 

Table~\ref{tab:benchresults} reports the alignment between model-inferred signatures and manually reconstructed reference headers, comparing our fine-tuned tool-calling \model (\textsc{Qwen3-8B}) with both static analysis and proprietary LLM APIs across numerical and structural inference metrics.

\subsection{Integrated Evaluation and Analysis}
\begin{figure*}[b!]
\centering

\begin{minipage}[t]{0.32\textwidth}
\centering
\begin{tikzpicture}
\begin{axis}[
    width=\linewidth,
    height=5.5cm,
    xlabel={\small Cost ($\downarrow$ better)},
    ylabel={\small Accuracy (\%) ($\uparrow$ better)},
    xmin=0, xmax=5,
    ymin=70, ymax=100,
    xtick={0,1,2,3,4,5},
    ytick={70,75,80,85,90,95,100},
    grid=both,
    legend style={at={(0.97,0.03)}, anchor=south east, font=\tiny, draw=black!20},
    legend cell align=left
]
\addplot+[only marks,mark=*,mark size=2pt,mark options={draw=black},fill=blue!30]
coordinates {(0.2,83.5)};
\addlegendentry{MOTIF-Model}
\addplot+[only marks,mark=square*,mark size=2pt,mark options={draw=black},fill=red!30]
coordinates {(5.0,86.3)};
\addlegendentry{GPT-4o}
\addplot+[only marks,mark=diamond*,mark size=2pt,mark options={draw=black},fill=green!30]
coordinates {(4.5,86.7)};
\addlegendentry{Claude 4}
\addplot+[only marks,mark=triangle*,mark size=2pt,mark options={draw=black},fill=purple!30]
coordinates {(1.0,82.9)};
\addlegendentry{Gemini 2.0}
\addplot+[only marks,mark=oplus*,mark size=2pt,mark options={draw=black},fill=orange!30]
coordinates {(0.8,79.5)};
\addlegendentry{DeepSeek R1}
\end{axis}
\end{tikzpicture}
\caption{Accuracy vs.\ cost on MOTIF.}
\label{fig:motif-cost}
\end{minipage}
\hfill
\begin{minipage}[t]{0.32\textwidth}
\centering
\begin{tikzpicture}
\begin{axis}[
    ybar,
    bar width=0.3em,
    width=\linewidth,
    height=5.5cm,
    ylabel={\scriptsize Metric Score (\%)},
    symbolic x coords={PM Acc, EM Acc, Tool Usage, Stability, Tool Corr, Tool HR},
    xtick=data,
    xticklabel style={rotate=45, anchor=east, font=\tiny},
    enlarge x limits=0.15,
    ymin=0, ymax=100,
    legend style={at={(0.5,1.05)}, anchor=south, legend columns=2, font=\tiny, draw=none, fill=none}
]
\addplot+[fill=gray!30] coordinates {(PM Acc,14.9) (EM Acc,0) (Tool Usage,0) (Stability,0) (Tool Corr,0) (Tool HR,0)};
\addplot+[fill=blue!30] coordinates {(PM Acc,71) (EM Acc,52) (Tool Usage,0) (Stability,0) (Tool Corr,0) (Tool HR,0)};
\addplot+[fill=green!30] coordinates {(PM Acc,83) (EM Acc,67) (Tool Usage,88) (Stability,20) (Tool Corr,100) (Tool HR,0)};
\addplot+[fill=orange!40] coordinates {(PM Acc,82) (EM Acc,61) (Tool Usage,99) (Stability,13) (Tool Corr,100) (Tool HR,0)};
\legend{Static, Zero-Shot, LLM+MOTIF, Ours+MOTIF}
\end{axis}
\end{tikzpicture}
\caption{Scalar and behavioral metrics.}
\label{fig:metric-num}
\end{minipage}
\hfill
\begin{minipage}[t]{0.32\textwidth}
\centering
\begin{tikzpicture}
\begin{axis}[
    ybar,
    bar width=0.3em,
    width=\linewidth,
    height=5.5cm,
    ylabel={\scriptsize Accuracy (\%)},
    symbolic x coords={BT Comp, Coll Inf, DP Inf, BT Inf},
    xtick=data,
    xticklabel style={rotate=45, anchor=east, font=\tiny},
    enlarge x limits=0.15,
    ymin=0, ymax=100,
    legend style={at={(0.5,1.05)}, anchor=south, legend columns=2, font=\tiny, draw=none, fill=none}
]
\addplot+[fill=gray!30] coordinates {(BT Comp,98) (Coll Inf,0) (DP Inf,0) (BT Inf,0)};
\addplot+[fill=blue!30] coordinates {(BT Comp,93) (Coll Inf,51) (DP Inf,52) (BT Inf,53)};
\addplot+[fill=green!30] coordinates {(BT Comp,96) (Coll Inf,59) (DP Inf,60) (BT Inf,60)};
\addplot+[fill=orange!40] coordinates {(BT Comp,96) (Coll Inf,33) (DP Inf,7) (BT Inf,24)};

\legend{Static, Zero-Shot, LLM+MOTIF, Ours+MOTIF}
\end{axis}
\end{tikzpicture}
\caption{Subtask-level inference accuracy.}
\label{fig:metric-sub}
\end{minipage}

\end{figure*}
\begin{figure*}[b!]
\centering
\begin{minipage}[t]{0.48\textwidth}
\centering
\begin{tikzpicture}
\begin{axis}[
    width=\linewidth,
    height=6cm,
    ybar,
    bar width=10pt,
    ylabel={Count},
    symbolic x coords={Conventional Types, Generic Collections, No ID Generics, No Structs, Selector Mismatch, Struct Refs, Method Not Parsed},
    xtick=data,
    xticklabel style={rotate=35, anchor=east, font=\small},
    nodes near coords,
    nodes near coords align={vertical},
    ymin=0,
    ymajorgrids=true,
    grid style=dashed,
    tick label style={font=\small, color=black},
    label style={font=\small, color=black},
    every axis plot/.append style={draw=black},
    every node near coord/.append style={font=\footnotesize, color=black},
    axis line style={black},
]
\addplot+[fill=green!20, draw=black] coordinates {
    (Conventional Types,1740)
    (Generic Collections,1163)
    (No ID Generics,569)
    (No Structs,186)
    (Selector Mismatch,366)
    (Struct Refs,198)
    (Method Not Parsed,480)
};
\end{axis}
\end{tikzpicture}
\caption{Empirical distribution of triggered constraints.}
\label{fig:linter-distribution}
\end{minipage}
\hfill
\begin{minipage}[t]{0.48\textwidth}
\centering
\begin{tikzpicture}
\begin{axis}[
    ybar,
    bar width=0.45em,
    width=\linewidth,
    height=6cm,
    ylabel={Median Gain (\%)},
    symbolic x coords={BT Completion, Collection Inference, DP Inference, BT Inference},
    xtick=data,
    xticklabel style={rotate=45, anchor=east},
    nodes near coords,
    ymin=0, ymax=30,
    enlarge x limits=0.2,
    axis y line*=left,
    axis x line*=bottom,
    legend style={at={(0.02,0.98)}, anchor=north west}
]
\addplot+[fill=blue!30] coordinates {
  (BT Completion,0.2)
  (Collection Inference,15.7)
  (DP Inference,26.4)
  (BT Inference,12.7)
};
\end{axis}
\end{tikzpicture}
\caption{Accuracy deltas per subtask.}
\label{fig:tool-ablation}
\end{minipage}
\hfill
\end{figure*}

\textbf{Unified Performance Comparison.}
We first do performance comparison across the full evaluation suite shown in Table~\ref{tab:motifbenchresults}, covering all four system configurations: static analysis, zero-shot LLM, tool-augmented LLM, and our fine-tuned 8B variant. 
Metrics are grouped into \textsc{(I)} scalar behavioural indicators (\eg, tool-usage efficiency, inference stability) and \textsc{(II)} structured subtask-level accuracy measures for signature inference. 

As shown in Figures~\ref{fig:metric-num} and~\ref{fig:metric-sub}, static analysis achieves 
low accuracy across all axes, recovering fewer than~2\% of valid structural signatures. 
Zero-shot LLMs improve moderately (up to~+18\,pp on PM accuracy) but remain brittle on type-inference and tool-alignment subtasks, with variance exceeding~0.25 across repeated runs. 
Introducing deterministic tool augmentation stabilises behaviour and increases overall signature recovery by~1.6$\times$. 

Ablation results in Table~\ref{tab:oneshot-vs-feedback} further isolate the effect of iterative feedback, showing that feedback-conditioned prompting yields an additional~12–15\,pp improvement in ambiguous-type precision and halves cross-run variance. 
This gain arises from grounding the model’s reasoning chain in verifiable tool outputs rather than unconstrained textual inference. 
Overall, the fine-tuned 8B configuration achieves parity with substantially larger proprietary LLMs while operating at one-eighth their parameter scale, reducing inference cost by over~60\% and improving controllability through consistent tool-bounded reasoning (see Figure~\ref{fig:motif-cost}). 

\textbf{Failure Analysis and Limitations.}
Figure~\ref{fig:linter-distribution} summarises the empirical distribution of triggered constraint categories derived from linter diagnostics.
The majority of violations arise from \emph{conventional type} and \emph{generic collection} inconsistencies (1{,}740 and 1{,}163 cases, respectively), together accounting for more than half of all detected issues.
These correspond to incomplete or missing generic annotations in Objective-C container types (\eg, \texttt{NSDictionary<NSString*,id>}), confirming that ambiguity in parameterised structures remains the dominant failure mode.
By contrast, structural mismatches, unresolved selectors, and unparsed method bodies contribute less than 15\% of the total, indicating that syntactic correctness is largely stabilised by the tool-bounded parsing stage.
Residual errors primarily stem from information not recoverable through static metadata alone (\eg, runtime type casts, opaque structs, or heavily optimised compiler output) representing intrinsic limitations of static-plus-LLM inference.

\textbf{Subtask-Level Improvements.}
Figure~\ref{fig:tool-ablation} quantifies median accuracy gains obtained through iterative feedback across the benchmark subtasks.
The most substantial improvement appears in \textsc{DP Inference} (+26.4 pp), followed by \textsc{Collection Inference} (+15.7 pp) and \textsc{BT Inference} (+12.7 pp), while \textsc{BT Completion} remains nearly unchanged (+0.2 pp).
These trends suggest that feedback-conditioned prompting most effectively enhances reasoning over high-entropy semantic spaces (
particularly data-path and collection-type inference) where raw textual priors are insufficient.
The relatively flat gain in boilerplate completion implies diminishing returns once syntactic structure is fully captured.
Together, 
they explain the aggregate improvements reported in Figure~\ref{fig:metric-sub} and Table~\ref{tab:oneshot-vs-feedback}, showing that most accuracy gains originate from enforcing and resolving type-generic constraints within the iterative loop.

\begin{table*}[!t]
\scriptsize
\caption{Ablation results showing the impact of tool augmentation and feedback loops on type inference quality.}

\setlength{\tabcolsep}{6pt}
\renewcommand{\arraystretch}{1.3}
\begin{tabular}{p{5cm} p{3cm} p{3.5cm} p{3cm}}
\toprule
\textbf{Setting} & \cellcolor{gray!10}\textbf{Exact Match Accuracy} & \cellcolor{orange!10}\textbf{Tool-Call Correctness} & \cellcolor{yellow!10}\textbf{Inference Stability} \\
\midrule

\rowcolor{catrow}
\multicolumn{4}{l}{\textit{Unassisted Inference (Baseline)}} \\
LLM (one-shot) & 23.1\% & -- & 42.0\% \\
\midrule
\rowcolor{catrow}
\multicolumn{4}{l}{\textit{Tool-Augmented Inference (No Feedback)}} \\
LLM + Tool Context & 44.7\% & 53.0\% & 55.8\% \\
\midrule
\rowcolor{catrow}
\multicolumn{4}{l}{\textit{Tool-Augmented Inference (With Feedback Loop)}} \\
\textbf{LLM + Tool + Feedback (Ours)} & \textbf{67.5\%} & \textbf{71.2\%} & \textbf{86.9\%} \\

\bottomrule
\end{tabular}
\label{tab:oneshot-vs-feedback}
\end{table*}
\section{Conclusion}

We introduced \motif, which encompasses a unified framework, benchmark, and model for automated reverse engineering of private Objective-C frameworks. Together, these components establish both a methodology and a resource for advancing automated reverse engineering.  Moreover, the design generalizes: by adapting tooling and prompt schemas, the same ReAct–feedback loop can be extended to other binary formats and operating systems.

\clearpage
\section*{Ethical Consideration}
Our work operates in a sensitive realm of reverse engineering private parts of a proprietary operating system. Our research is conducted strictly under the banner of academic exploration and security research, not for building software intended for consumer distribution or retail within Apple’s ecosystem. To ensure ethical research,
we do not distribute private framework binaries or reconstructed APIs in application packages intended for external deployment. Further, our tools and models are shared only for research purposes, not embedded in consumer-facing software, especially not via the App Store. By framing MOTIF as a research platform rather than a production tool, we aim to improve transparency and robustness of closed-system platforms while remaining mindful of legal and ethical constraints.



{\footnotesize \bibliographystyle{acm}
\bibliography{sample}}

\appendix
\subsection{Case Study: \agent System Trace for Recovering Hidden API Signatures}
\label{sec:appendix_example}

To illustrate the internal operation of the \agent, we present a concrete end-to-end trace on a representative reverse-engineering task recovering the missing Objective-C method signature for a private macOS framework header. Table \ref{tab:case-meta} summarises the key experimental context, including the target operating system version, framework, and header file under analysis.

\begin{table}[h!]
\caption{Metadata for case study. The listed configuration corresponds to the runtime setup used in Figures \ref{fig:motif_case_study1}, \ref{fig:motif_case_study2}, \ref{fig:motif_case_study3}.}
\centering
\label{tab:case-meta}
\begin{tabular}{l l}
\toprule
\textbf{Attribute} & \textbf{Value} \\
\midrule
Target OS version & macOS 14.5 (Sonoma) \\
Target framework & \texttt{IOUSBHost.framework} \\
Target header & \texttt{IOUSBHostObject.h} \\
Analysis date & August 2025 \\
MOTIF agent version & v1.3 (default configuration) \\
LLM backend & \texttt{GPT-4o-mini} \\
Model size & $\sim$38B parameters \\
Context window & 128k tokens \\
Candidate pool size & 5 (top-1 correct) \\
\bottomrule
\end{tabular}
\end{table}

This case study serves as a compact yet informative example, demonstrating how the agent orchestrates reasoning, tool invocation, and constraint-guided validation to infer precise type semantics.

\subsection{Problem Setup and Symbolic Domains}
\label{sec:formalization}
\textbf{Header corpus.}
We begin with a Mach-O binary with partially stripped Objective-C interface metadata. Using compiler-level introspection, we statically extract headers $\mathcal{H}=\{h_1,\dots,h_m\}$ and parse each $h_i$ under a deterministic LL(*) grammar $\mathbb{G}_{\text{ObjC}}$ into an abstract syntax tree (AST), forming $\mathcal{A}=\bigcup_i \texttt{AST}_{h_i}$.

\textbf{Ambiguity set.}
Within $\mathcal{A}$ we target underspecified declarations $\mathcal{G}\subset\mathcal{A}$ whose return or parameter types belong to the finite ambiguous set $\mathcal{T}_{\text{ambig}}=\{\texttt{id},\texttt{void *},\texttt{struct\{...\}}, \dots\}$. Resolving these ambiguities requires mapping syntactic declarations to their compiled representations.

\textbf{Symbol binding.}
For every $g_i\in\mathcal{G}$ we establish a static correspondence to a unique binary selector $s_i\in\mathcal{S}$ extracted from the Mach-O symbol table. Selectors encode both class and method components (e.g., \texttt{+[CBUUID UUIDWithData:]}), providing a bijective mapping between AST-level declarations and compiled symbols. This mapping anchors source-level ambiguity to a concrete binary artefact and enables cross-referencing of type hypotheses against compiled metadata.

\textbf{Inference context (LLM input).}
For each recovery target $g_i$ we assemble the prompt context $\mathcal{T}_i=(\mathcal{N}_i,\mathcal{I}_i)$ that \emph{excludes} semantic diagnostics. $\mathcal{N}_i$ captures natural source evidence (raw declaration text, lexical neighbourhood, identifier and file locality). $\mathcal{I}_i$ aggregates symbolic descriptors available to tools (disassembly and decompilation at $s_i$, typedef resolution in $\mathcal{A}$, and symbol metadata/addresses). The template $P_i$ instantiated with $\mathcal{T}_i$ yields an initial fully-specified candidate signature $\hat{g}$.

\textbf{Constraint formulation and refinement loop.}
Our goal is to identify the most semantically admissible signature consistent with all high-severity constraints. Given $\hat{g}$, a stratified linter $\mathcal{L}\!:\!\texttt{Signature}\!\rightarrow\!\mathcal{M}$ produces a multiset of messages $\mathcal{M}=\{(m_j,s_j)\}_{j=1}^{k}$ with severities $s_j\in\{\texttt{low},\texttt{medium},\texttt{high}\}$. Messages induce Boolean constraints $c_j$ partitioned as
$\mathcal{C}_{\texttt{hard}}=\{c_j\mid s_j=\texttt{high}\}$ and
$\mathcal{C}_{\texttt{soft}}=\{c_j\mid s_j\in\{\texttt{medium},\texttt{low}\}\}$.
We pose selection as a \emph{Weighted Partial MaxSAT} objective over $\texttt{Signature}$:
\begin{align}
\hat{g}^* &= 
\arg\min_{\hat{g}}
\sum_{c_j \in \mathcal{C}_{\texttt{soft}}}
\lambda(s_j)\,\mathbb{1}[c_j(\hat{g})=\texttt{false}],
\\[2pt]
&\text{s.t. }\forall c_j\in\mathcal{C}_{\texttt{hard}},\;
c_j(\hat{g})=\texttt{true}.
\end{align}
Here $\lambda(\cdot)$ weights soft violations. If $\mathcal{M}\neq\emptyset$, the structured feedback is fed back to re-instantiations of $P_i$ (same $\mathcal{T}_i$, updated guidance), iterating until all hard constraints are satisfied and the soft cost stabilises. This separation keeps LLM inputs strictly evidential ($\mathcal{N}_i,\mathcal{I}_i$) while delegating admissibility to the constraint layer.

\begin{figure*}[h!]
  \centering
  \includegraphics[width=\linewidth]{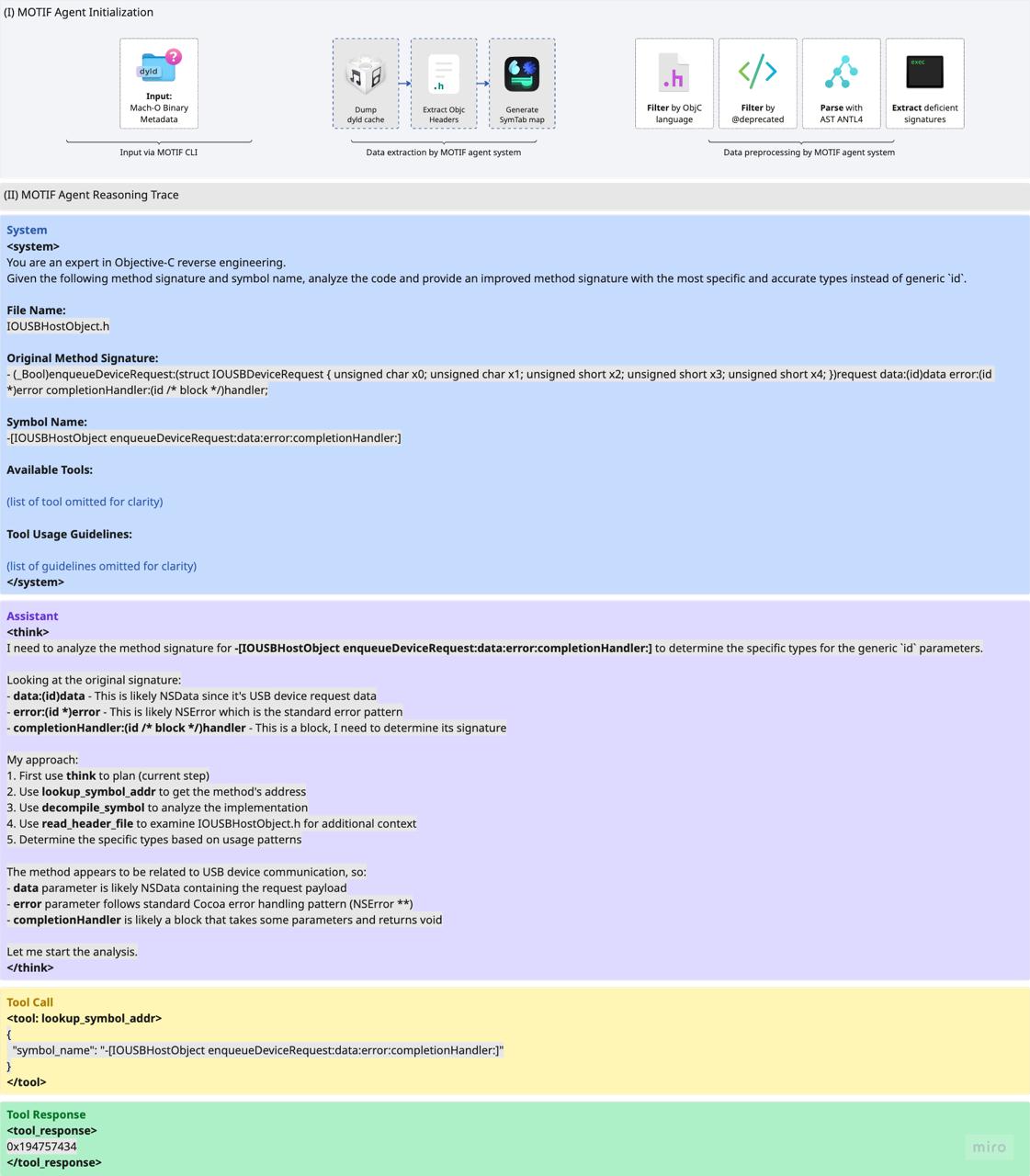}
  \caption{
  Step-wise execution of the \textbf{MOTIF agent} on a representative case.
  (I) \textbf{Initialization:} the agent loads Mach-O metadata, configures tool interfaces, and prepares the reasoning environment.
  (II) \textbf{Reasoning Trace:} the agent performs symbolic analysis, tool calls, and decompilation to infer type semantics.
  (III) \textbf{Constraint Satisfaction and Linter Feedback:} candidate signatures are evaluated under the SAT\_INFER objective;
  the optimal candidate $\hat{g}^*$ is selected with zero hard-constraint violations and minimal soft-constraint cost.
  If no valid candidate exists, the linter feedback triggers a new reasoning iteration.
  }
  \label{fig:motif_case_study1}
\end{figure*}

\begin{figure*}[h!]
  \centering
  \includegraphics[width=\linewidth]{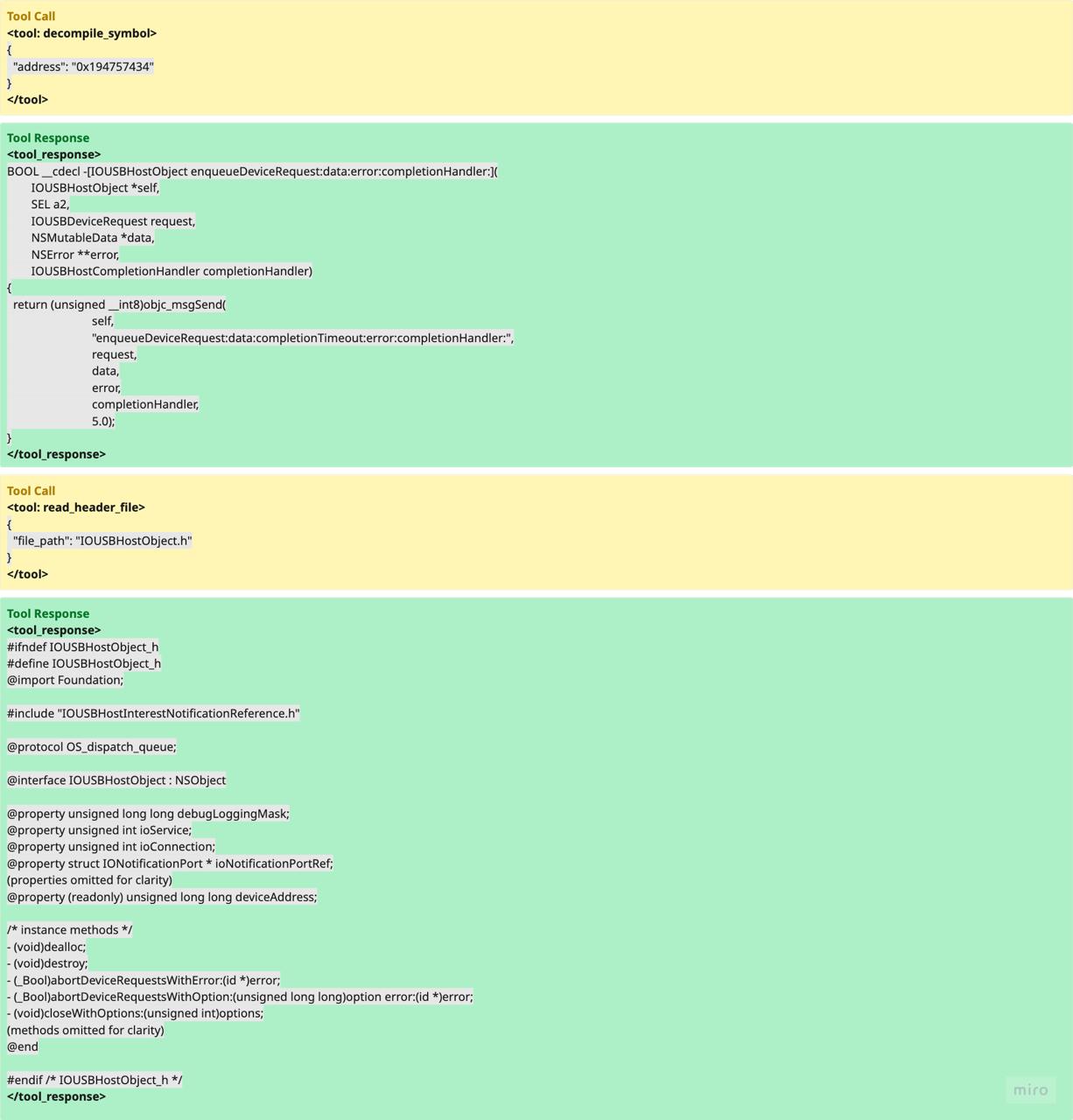}
  \caption{
  Step-wise execution of the \textbf{MOTIF agent} on a representative case.
  (I) \textbf{Initialization:} the agent loads Mach-O metadata, configures tool interfaces, and prepares the reasoning environment.
  (II) \textbf{Reasoning Trace:} the agent performs symbolic analysis, tool calls, and decompilation to infer type semantics.
  (III) \textbf{Constraint Satisfaction and Linter Feedback:} candidate signatures are evaluated under the SAT\_INFER objective;
  the optimal candidate $\hat{g}^*$ is selected with zero hard-constraint violations and minimal soft-constraint cost.
  If no valid candidate exists, the linter feedback triggers a new reasoning iteration.
  }
  \label{fig:motif_case_study2}
\end{figure*}

\begin{figure*}[h!]
  \includegraphics[width=\linewidth]{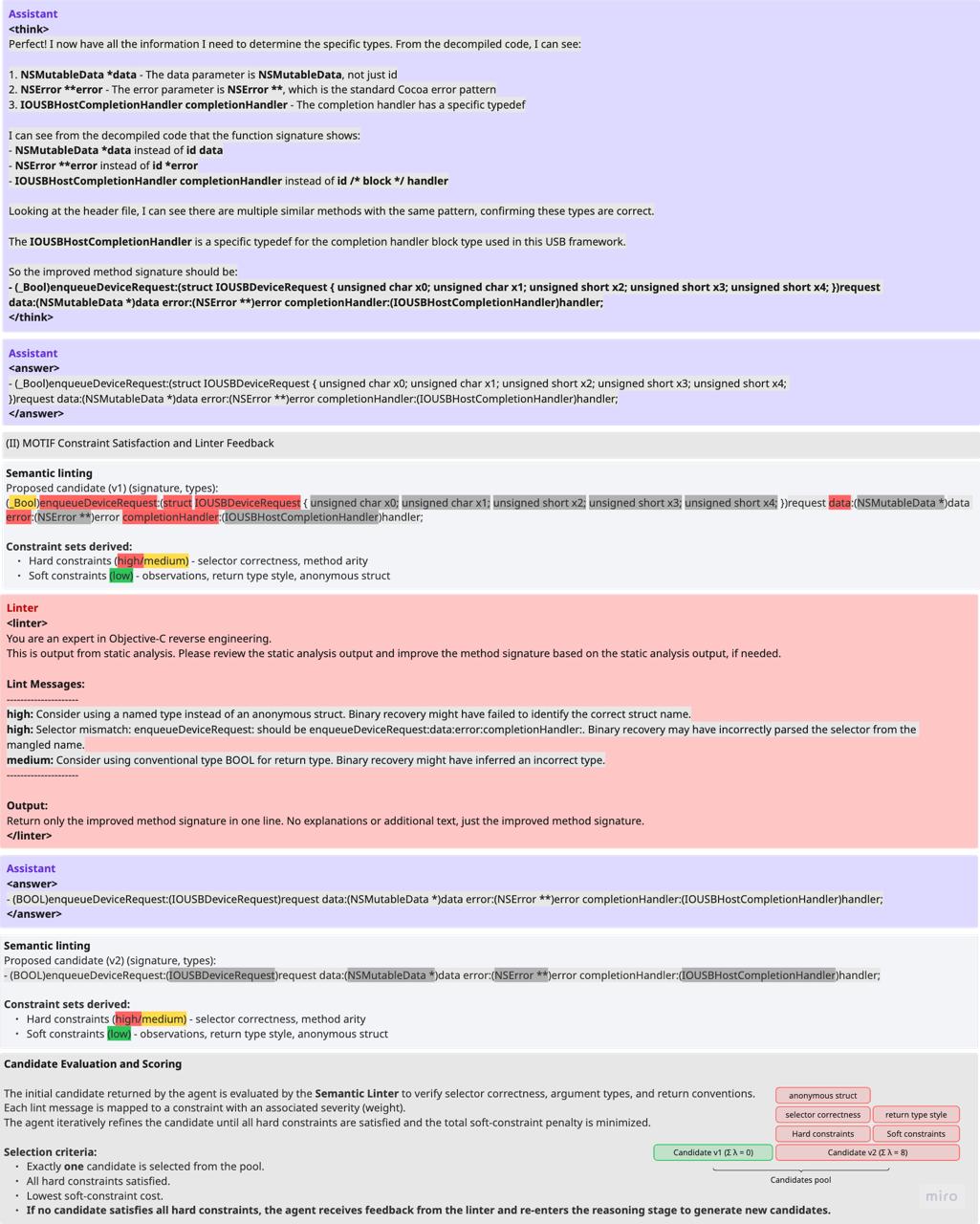}
  \caption{
  Step-wise execution of the \textbf{MOTIF agent} on a representative case.
  (I) \textbf{Initialization:} the agent loads Mach-O metadata, configures tool interfaces, and prepares the reasoning environment.
  (II) \textbf{Reasoning Trace:} the agent performs symbolic analysis, tool calls, and decompilation to infer type semantics.
  (III) \textbf{Constraint Satisfaction and Linter Feedback:} candidate signatures are evaluated under the SAT\_INFER objective;
  the optimal candidate $\hat{g}^*$ is selected with zero hard-constraint violations and minimal soft-constraint cost.
  If no valid candidate exists, the linter feedback triggers a new reasoning iteration.
  }
  \label{fig:motif_case_study3}
\end{figure*}

\end{document}